\documentstyle[epsf,aps]{revtex}
\newcommand{\beq}{\begin{equation}}
\newcommand{\eeq}{\end{equation}}
\newcommand{\beqa}{\begin{eqnarray}}
\newcommand{\eeqa}{\end{eqnarray}}

\newcommand{\barr}[1]{\not\mathrel #1}
\newcommand{\vs}{\vspace{-0.0cm}}

\begin{document}
\title{
\begin{flushright}
{\normalsize{FZJ-IKP(TH)-2000-02}}\\
{\normalsize{}}
\end{flushright}
\vspace{2 cm}
Ordinary and radiative muon capture on the proton\\[0.2em]
and the pseudoscalar form factor of the nucleon}
\vspace{2.5 cm}
\author{V{\' e}ronique Bernard$^a$\footnote{email: bernard@lpt6.u-strasbg.fr },
Thomas R. Hemmert$^b$\footnote{email: th.hemmert@fz-juelich.de}
and Ulf-G. Mei{\ss}ner$^b$\footnote{email: Ulf-G.Meissner@fz-juelich.de} \\
[1cm]
{\small {$^a$ Universit\'e Louis Pasteur, Laboratoire de Physique
Th\'eorique, F-67084 Strasbourg, France}} \\
{\small {$^b$ Forschungszentrum J{\" u}lich, Institut f{\" u}r Kernphysik
(Th), D-52425
J{\" u}lich, Germany}}}
\maketitle

\thispagestyle{empty}

\vspace{2cm}

\begin{abstract}
We calculate ordinary and radiative muon capture on the proton
in an effective field theory of pions, nucleons and delta isobars, working to
third and second order in the small scale expansion respectively. Preceding
calculations in chiral effective field theories only employed pion and
nucleon degrees of freedom and were not able to reproduce the photon spectrum
in the
pioneering
experiment of radiative muon capture on the proton from TRIUMF. For the past
few
years it has been speculated that the discrepancy between theory and experiment
might be due to $\Delta$(1232) related effects, which are only included via
higher
order contact interactions in the standard chiral approach. In this report
we demonstrate that this speculation does not hold true. We show that contrary
to
expectations from naive dimensional analysis isobar effects on the photon
spectrum
and the total rate in radiative muon capture are of the order of a few percent,
consistent with earlier findings in a more phenomenological approach. We
further
demonstrate that both ordinary and radiative muon capture constitute systems
with a
very well behaved chiral expansion, both in standard chiral perturbation theory
and in
the small scale expansion, and present some new ideas that might
be
at the bottom of the still unresolved
discrepancy between theory and experiment in radiative muon capture.
Finally we comment upon the procedure employed by the TRIUMF group to extract
new
information from their radiative muon capture experiment on the pseudoscalar
form factor
of the nucleon. We show that it is inconsistent with the ordinary muon
capture data.

\vspace{1cm}

\noindent PACS: 23.40.-s,12.39.Fe,13.60-r

\noindent Keywords: Ordinary and radiative muon capture, pseudoscalar form
factor,
chiral effective field theory
\end{abstract}
\newpage

%%%%%%%%%%%%%%%%%%%%%%%%%%%%%%%%%%%%%%%%%%%%%%%%%%%%%%%%%%%%%%%%%%%%%%%%%%%%%%%%%%%
\section{Introduction}
\label{sec:insum}
\def\theequation{\arabic{section}.\arabic{equation}}
\setcounter{equation}{0}
\setcounter{page}{1}

With the ubiquitous success of the Standard Model\footnote{For a recent review
of
utilizing various muon capture reactions as filters for physics beyond the
Standard
Model see \cite{GL00}.} ordinary and radiative
muon capture on the proton can nowadays be
considered as an excellent testing ground for our understanding of
spontaneous and explicit chiral symmetry breaking in QCD. This stems
from the fact that the typical momentum transfer in these reactions
is very small---of the order of the muon mass---and one therefore can
apply effective field theory methods, in particular baryon chiral
perturbation theory.

Ordinary muon capture (OMC),
\beq\label{OMCdef}
\mu^-(l)+p(r)\rightarrow \nu_\mu(l^\prime) +n(r^\prime)~,
\eeq
where we have indicated the four--momenta of the various particles,
allows to measure the so--called induced pseudoscalar coupling
constant, $g_P$. This coupling constant is nothing but the value
of the induced pseudoscalar form factor $G_P (t)$ at the four--momentum
transfer
for muon capture by the proton at rest, $g_p = m_\mu \, G_P
(t=-0.88m^2_\mu) / 2M_N$, with $m_\mu \, (M_N)$ the muon  (nucleon) mass.
Theoretically, it is dominated by the pion
pole as given by the time--honored PCAC prediction,
$g_P^{\rm PCAC} = 8.89$. Adler and Dothan~\cite{AD}
as well as Wolfenstein~\cite{Wo} calculated the first correction to the
PCAC result utilizing by now outdated (and in some cases misleading)
methods. The ADW relation was rederived directly from QCD Ward
identities within the framework of heavy baryon chiral perturbation
theory~\cite{bkmgp} and shown not to be affected by $\Delta (1232)$
effects in~\cite{BFHM}, $g_P^{\rm CHPT} = 8.44\pm 0.23$ (a similar calculation
with a slightly different result was given
in~\cite{FLMS}). The presently available data have, however, too
large error bars to discriminate between the pion pole prediction and
its corrected version~\cite{bald}. Furthermore, $G_p (t)$ for low
four--momentum
transfer squared can be extracted from charged pion electroproduction
measurements~\cite{choi}. The resulting momentum dependence of the
induced pseudoscalar form factor is in agreement with the pion pole
prediction. However, as in the case of the pseudoscalar coupling,
the data are not precise enough to be sensitive
to the small corrections found in~\cite{AD}\cite{Wo}\cite{bkmgp}.

While the momentum transfer in OMC is fixed, radiative
muon capture (RMC),
\beq\label{RMCdef}
\mu^-(l)+p(r)\rightarrow \nu_\mu(l^\prime) +n(r^\prime)+\gamma(k)~,
\eeq
has a variable momentum transfer $t$ and one can get up to $t = m_\mu^2$
at the maximum
photon energy of about $k \sim 100\,$MeV, which is quite close to the
pion pole. This amounts approximately to a four times larger
sensitivity to $g_P$ in RMC than OMC. However, this increased
sensitivity is upset by the very small partial branching ratio in hydrogen
$(\sim 10^{-8}$ for photons with $k > 60\,$MeV) and one thus has to
deal with large backgrounds.
Precisely for this reason only very recently a first measurement
of RMC on the proton has been published~\cite{TRIUMF}. The resulting
number for $g_P$, which was obtained using a relativistic tree model
including the $\Delta$--isobar~\cite{BF2}, came out significantly
larger than expected from OMC, $g_P^{\rm RMC} = 12.35 \pm 0.88 \pm
0.38 = 1.46 \, g_P^{\rm CHPT}$. It should be noted that in this model the
momentum
dependence in $G_P(t)$ is solely given in terms of the pion pole and the
induced
pseudoscalar coupling is obtained as a multiplicative factor from
direct comparison to the photon spectrum and the partial RMC branching
ratio (for photon energies larger than 60 MeV). It was also argued
in~\cite{TRIUMF} that the atomic and molecular physics related to the
binding of the muon in singlet and triplet atomic $\mu p$ and ortho
and para $p\mu p$ molecular states is sufficiently well under control.

The TRIUMF result spurred a lot of theoretical activity. While radiative
muon capture had already been calculated in phenomenological tree
level models a long time ago,
see e.g.~\cite{Opat}\cite{Beder}\cite{Fear}\cite{BF1}\cite{BF2},
heavy baryon chiral perturbation theory was also used at tree level including
dimension
two operators~\cite{meiss} and to one loop order~\cite{AM}. The
resulting photon spectra are not very different from the ones obtained
in the phenomenological models, the most striking feature being the
smallness of the chiral loops~\cite{AM}, hinting towards a good convergence of
the chiral expansion. At present, the puzzling result from the TRIUMF
experiment remains unexplained. It is, however, a viable possibility
that the discrepancy does not come from the strong interactions but
rather is related to the distribution of the various spin states of the
muonic atoms. It is therefore mandatory to sharpen the theoretical
predictions for the strong as well as the non--strong physics entering
the experimental analysis.\footnote{A recently proposed solution to
  the problem~\cite{Che} based on a novel term has been shown to be
  inconsistent with CVC in ref.~\cite{fearc}. The
  corrected term--which in the chiral counting comes in at ${\cal
O}(p^3)$--only
  gives a small contribution and is already
  contained in the RMC calculation of Ando and Min \cite{AM}.}

Here, we wish to reanalyze RMC in the framework
of the so--called small scale expansion~\cite{hhkl,hhk}, which allows to
systematically include the $\Delta$ resonance into the effective field theory.
Although Ando and Min \cite{AM} have already shown that the RMC process
possesses a well behaved
chiral expansion up to N$^2$LO, it has been noted quite
early \cite{Nimai} that one should reanalyze
RMC in a chiral effective field theory with explicit $\Delta$ degrees of
freedom\footnote{Phenomenological models have claimed for a long time that the
$\Delta$ contribution does not exceed 8\% in the photon spectrum for photon
energies
above 60~MeV~\cite{BF1}.}. This
is due to the fact that the $\Delta$-resonance lies quite close to the nucleon
and
therefore, in a delta-free theory like HBChPT as used in \cite{AM}, could lead
to
unnaturally
large higher order contact interactions which would spoil the seemingly good
chiral
convergence. Stating the same concern in the language of (naive) dimensional
analysis,
it suggests the possibility of corrections of the order
of 30\% due to the small nucleon--delta mass splitting, $m_\mu
/(M_\Delta - M_N) \sim 3 m_\mu / M_N$. In the small scale expansion,
the leading delta effects involving the large M1 $\gamma N\Delta$ vertex
already appear
at second order $\epsilon^2$ ($\epsilon$ denotes a genuine small parameter,
the pion mass, external momentum or the $N\Delta$ mass splitting) and can
therefore
already be analyzed at tree level.
The study presented here therefore constitutes a natural extension of the two
previous
analyses of RMC \cite{meiss,AM} using chiral effective field
theories\footnote{At
the moment
there exists no systematic scheme in chiral effective field theories to
simultaneously
include the effects of explicit vector mesons and other strong short range
effects into
the calculation. Given the small momentum transfer of RMC in the t-channel one
expects,
however, that these contributions can be correctly accounted for via ${\cal
O}(p^3)$
contact interactions \cite{AM}. A Born term analysis of vector meson
contributions in the
RMC process has been presented in \cite{STK}.}.
Some preliminary results where reported in ref.\cite{B98}.
As we will show, the small scale expansion allows for a very
transparent separation of the resonance and chiral pion effects.

The manuscript is organized as follows. In section~2, we briefly review how
the Standard Model at low energies is mapped onto a chiral effective field
theory. The ingredients of this field theory, which uses  pions, nucleons
and the delta isobar as degrees of freedom, are discussed in section~3.
Section~4 is concerned with ordinary and radiative muon capture on the proton.
In the
framework of the small scale expansion in OMC
the leading order delta effects only come in at N$^2$LO ({\it i.e.}
${\cal O}(\epsilon^3)$), whereas for RMC one can already study the explicit
influence
of this resonance at NLO ({\it i.e.} ${\cal O}(\epsilon^2)$).
The status of the induced pseudoscalar
form factor is reviewed and the analysis of the TRIUMF RMC experiment
is critically assessed in section~5. Section~6 contains the summary and
conclusions. Some more technical aspects of this work are relegated to the
appendix.

%%%%%%%%%%%%%%%%%%%%%%%%%%%%%%%%%%%%%%%%%%%%%%%%%%%%%%%%%%%%%%%%%%%%%%%%%%%%%%%
\section{From the Standard Model to the effective field theory}
\label{sec:EFT}
\def\theequation{\arabic{section}.\arabic{equation}}
\setcounter{equation}{0}

In this section we briefly summarize how QCD coupled to the standard
electroweak theory is mapped onto the pertinent effective field theory.
For that, we start with the electroweak Lagrangian. For our purpose, we only
need the coupling of the charged currents to the charged massive
gauge bosons $(W^\pm_\mu$) and the coupling of the massless photon
($A_\mu$) to the electromagnetic (em) current,
\beq
{\cal L}_{\rm int}^{\rm SM}= -\frac{g_2}{\sqrt{8}}
\left\{ W_\mu^+(x) J_{\rm ch}^{\mu}
+W_\mu^-(x)
J_{\rm ch}^{\mu\dagger} \right\} -g_1\cos\theta_W A_\mu(x)
J^\mu_{\rm em}+{\cal L}_{\rm ntl-wk}
\eeq
where the last term refers to the non--leptonic weak interactions.
$g_1$ and $g_2$ are the gauge couplings of the U(1)$_Y$ and the
SU(2)$_L$ gauge groups and $\theta_W$ is the weak mixing (Weinberg) angle.
In terms of the light quark and lepton fields the em and charged
weak currents read
\begin{eqnarray}
J_{\rm em}^\mu&=&\frac{2}{3}\bar{u}\gamma^\mu u-\frac{1}{3}\bar{d}\gamma^\mu d
             -\bar{\mu}\gamma^\mu \mu + \dots \\
J_{\rm ch}^\mu&=&\bar{u}_w\gamma^\mu\left(1-\gamma_5\right)d_w +
             \bar{\nu}_\mu \gamma^\mu\left(1-\gamma_5\right)\mu + \dots~,
\end{eqnarray}
with $q_w=(u_w,d_w)$ denoting the quark eigenstates of the weak interaction and
the ellipsis denoting terms we do not need in what follows.
For constructing the effective field theory, it is most convenient to
consider the QCD Lagrangian coupled to these gauge fields treated as
external local sources, which transform locally under chiral
symmetry~\cite{GL}.
With $e=g_1\cos\theta_W$ and $W_\mu^-={\cal V}_\mu^-(x)-{\cal A}_\mu^-(x)$,
QCD coupled to these external sources takes the form
\begin{eqnarray}\label{LQCDex}
{\cal L}_{\rm QCD}^{\rm ext. fields}&=&{\cal L}_{\rm
QCD}^0+\bar{q}\left[\not{\bf
v}(x)-\not{\bf a}(x)
\gamma_5\right]q-\bar{q}\left[{\bf s}(x)-i\,{\bf p}(x)\right]q \nonumber \\
& &+\bar{q}\left[\not{v}^{(0)}(x)-\not{a}^{(0)}(x)
\gamma_5\right]q-\bar{q}\left[s^{(0)}(x)-i\,p^{(0)}(x)\right]q~,
\end{eqnarray}
with $q$ the bi--spinor of the light quark fields in the strong interaction
basis.
For the case of RMC we are confronted with the following scenario of external
sources\footnote{We are working in the limit of no isospin breaking, i.e.
equal quark masses $m_u=m_d$ and no internal (virtual) photon effects.}:
\begin{eqnarray}
{\bf s}(x)&=&0~, \label{eq:exta}\\
s^{(0)}(x)&=&\hat{m}\, I~, \\
{\bf p}(x)&=&p^{(0)}(x)\;=\; 0~, \\
{\bf v}_\mu(x)&=&-e\,\frac{1}{2}\tau^3\,A_\mu(x)
                -\biggl[\frac{g_2 V_{ud}}{\sqrt{8}}\left(
                \frac{1}{2}\tau^1-\frac{i}{2}
                \,\tau^2\right)
                {\cal V}_\mu^-(x) + {\rm h.c.}\biggr]~, \\
v_\mu^{(0)}(x)&=&-e\,\frac{1}{6}I\,A_\mu(x)~, \\
{\bf a}_\mu(x)&=&-\frac{g_2
V_{ud}}{\sqrt{8}}\left(\frac{1}{2}\tau^1-\frac{i}{2}\,
                \tau^2\right)
                {\cal A}_\mu^-(x) + {\rm h.c.}~, \\
a_\mu^{(0)}&=&0 \label{eq:extz}~,
\end{eqnarray}
with $s(x),p(x),v_\mu (x)$ and $a_\mu (x)$ scalar, pseudoscalar, vector and
axial--vector fields, in order, in an obvious isospin (singlet and
triplet) notation.  $V_{ud}$ is the pertinent element of the
CKM matrix. Note that the term ${\cal L}_{\rm QCD}^0$ in
Eq.(\ref{LQCDex}) is chirally symmetric. The explicit chiral symmetry
breaking due to the current quark masses resides in the zeroth
component of the external scalar source.

Having specified the external field environment we now analyze the required
strong matrix elements for calculating OMC and RMC.
First we define quark vector and axial-vector currents
\begin{eqnarray}
V_\mu^a &=& \bar{q}\,\gamma_\mu \,T^a \,q~, \\
A_\mu^a &=& \bar{q}\, \gamma_\mu\gamma_5 \, T^a \,q \; .
\end{eqnarray}
Here, the $T^a = \tau^a/2$ are the generators of SU(2).
With these definitions one then specifies the strong matrix elements which have
to be calculated in the effective theory
\begin{eqnarray}
a)~& &\langle N|V_\mu^a |N\rangle~, \label{strongME1}\\
b)~& &\langle N|A_\mu^a |N\rangle~, \label{strongME2}\\
c)~& &\langle N|{\cal T} \,V_\mu^a V_\nu^b |N\rangle~, \label{strongME3}\\
d)~& &\langle N|{\cal T} \,V_\mu^a A_\nu^b |N\rangle \label{strongME4}\; ,
\end{eqnarray}
with all possible combinations of the external fields of
Eqs.(\ref{eq:exta}-\ref{eq:extz}) and ${\cal T}$ denotes
the conventional time--ordering operator. To proceed, we now have to
specify the effective Lagrangian which will be used to calculate
these matrix elements.

%%%%%%%%%%%%%%%%%%%%%%%%%%%%%%%%%%%%%%%%%%%%%%%%%%%%%%%%%%%%%%%%%%%%%%%%%%%%%%%%%%
\section{Chiral Lagrangians}
\label{sec:Lagr}
\def\theequation{\arabic{section}.\arabic{equation}}
\setcounter{equation}{0}

In this section, we briefly review the chiral Lagrangian underlying our
calculation. While the meson part is standard, with the external momenta
and the pion (light quark) mass counted as small parameters,
for the meson--baryon
system we include the nucleons as well as the $\Delta (1232)$ resonance.
To systematically account for the effects of the latter, the $N\Delta$ mass
splitting is considered as an additional small parameter. This is routed
in phenomenology and the large $N_c$ limit of QCD, but not in the strict chiral
limit, where the $\Delta$ decouples. These three small parameters are
collectively denoted by $\epsilon$.

\subsection{Meson chiral perturbation theory}

At low energies the chiral symmetry of QCD is spontaneously broken to the
vectorial subgroup $SU(2)_V$: $SU(2)_L \times SU(2)_R \rightarrow SU(2)_V$.
The strictures of the spontaneous and the explicit chiral symmetry breaking
can be explored in terms of an effective field theory, chiral perturbation
theory (ChPT). As a tool, one works with an effective Lagrangian, which
consists of a string of terms with increasing dimension. For our purpose,
we only need the first term in this expansion, the non-linear $\sigma$--model
chirally coupled to the external sources,
\begin{eqnarray}
{\cal L}_{\pi\pi}^{(2)}&=&\frac{F_{0}^2}{4}\,{\rm Tr}\left[\nabla_\mu U^\dagger
\nabla^\mu U
                          +\chi^\dagger U+\chi U^\dagger\right]~
\end{eqnarray}
with
\begin{eqnarray}
U(x)&=&\exp\left\{\frac{i}{F_0}\,\tau\cdot\pi(x)\right\}~, \\
\nabla_\mu U&=&\partial_\mu U-i({\bf v_\mu}+{\bf a_\mu})U+iU({\bf v_\mu}-
{\bf a_\mu})~, \\
\chi&=&2 B_0\left({\bf s}+s^{(0)}+i\,{\bf p}+i\,p^{(0)}\right)~.
\end{eqnarray}
The pions, which are nothing but integration variables, are collected in
the matrix--valued field $U(x)$ and the explicit symmetry breaking due to the
light quark masses is hidden in  $\chi$ via the zeroth component of the
scalar source. $F_0$ is the (weak) pion decay constant (in the chiral limit)
and $B_0$ is related to the scalar quark condensate. We work in
the standard scenario with $B_0 \gg F_0$. This specifies completely the
meson part of the effective Lagrangian.

\subsection{Including baryons: The small scale expansion}

The nucleon--delta--pion system chirally coupled to the external fields
can also be represented by a Lagrangian, which decomposes into a string
of terms with increasing dimension. We work here in the heavy mass
formulation, in which the nucleon and the delta are essentially considered
as heavy, static sources. This allows to shuffle the baryon mass ($M_B$)
dependence into  a string of $1/M_B$ suppressed vertices and gives rise to a
consistent power counting~\cite{jm}\cite{bkkm}\cite{hhk}.
Denoting by $N$ the (heavy) nucleon isodoublet field and by
$T_\mu^i$ the Rarita--Schwinger representation of the heavy
spin-3/2 field, the lowest order terms read
\begin{eqnarray}
{\cal L}_{\pi N}^{(1)} & = & \bar{N} \left[ i \; v \cdot D \; + \; \dot{g}_{A}
                             \; S \cdot u \right] N ~, \nonumber\\
{\cal L}_{\pi\Delta}^{(1)} & = & - \; \bar{T}^{\mu}_{i} \left[ i \; v\cdot
                            D^{ij}- \delta^{ij}\Delta_0 + \ldots
                            \right] \; g_{\mu\nu} \; T^{\nu}_{j}~,\nonumber\\
{\cal L}_{\pi N\Delta}^{(1)} & = & \dot{g}_{\pi N\Delta} \; \left\{
\bar{T}^{\mu}_{i}
                                  \; g_{\mu\alpha} \; w^{\alpha}_{i} \; N \;
                                  + \; \bar{N} \; w^{\alpha\dagger}_{i} \;
                                  g_{\alpha\mu} \; T^{\mu}_{i}
                                  \right\}~,   \label{eq:oo}
\end{eqnarray}
with $\Delta_0=M_{\Delta} - M_{0} \,(M_0)$ being the  nucleon-delta
mass splitting (nucleon mass) in the chiral limit (to the order we are
working, we can set $\Delta_0 = \Delta$). All other quantities $Q$ in
the chiral limit are
denoted as $\dot{Q}$.
Furthermore, $v_\mu$ and $S_\mu$ are the four--velocity and the
spin--vector of the heavy nucleon.  The various chiral covariant derivatives,
chiral connections and vielbeins are (for details, we refer to~\cite{bkmrev}
and \cite{hhk})
\begin{eqnarray}
D_\mu N&=&(\partial_\mu+\Gamma_\mu -i v_{\mu}^{(s)})N~,\nonumber \\
v_\mu^{(s)}&=&3\,v_\mu^{(0)}~, \nonumber \\
\Gamma_\mu&=&{1\over 2}[u^\dagger,\partial_\mu u]-{i\over
2}u^\dagger({\bf v_\mu}+{\bf a_\mu})u-{i\over 2}u({\bf v_\mu}-{\bf
a_\mu})u^\dagger\equiv \tau^i \; \Gamma_{\mu}^i~, \nonumber \\
u_\mu&=&iu^\dagger\nabla_\mu Uu^\dagger\equiv \tau^i \; w_{\mu}^i
\nonumber~, \\
D_\mu^{ij}T_\nu^j&=&\left(\partial_\mu\delta^{ij}+C_\mu^{ij}\right)T_\nu^j
\nonumber~, \\
C^{ij}_\mu&=&\delta^{ij}\left(\Gamma_\mu-i v^{(s)}_\mu\right)
-2i\epsilon^{ijk}\Gamma^k_\mu~,
\end{eqnarray}
where $i,j,k$ are isospin indices.
At second order in the small scale expansion, we need the following terms
\begin{eqnarray}
{\cal L}_{\pi N}^{(2)}&=&{1\over 2M_0}\bar{N}\left\{ (v\cdot D)^2-D^2
                   -ig_A\left(S\cdot D\,v\cdot u+v\cdot u\, S\cdot D\right)
                   \right. \\
                & &\phantom{{1\over 2M_0}\bar{N}}
                   \left. - \frac{i}{2} \left[ S^\mu , S^\nu \right]_- \left[
                   \left(1+\dot{\kappa}_v\right) f_{\mu\nu}^+ + 2 \left(1+
                   \dot{\kappa}_s\right) v_{\mu\nu}^{(s)} \right] +\dots
                   \right\} N~, \nonumber \\
{\cal L}_{\pi N\Delta}^{(2)}&=&\bar{T}^{\mu}_i \; \frac{1}{2M_0}
                           \left[ b_1 \; i f_{+\mu\nu}^{\; i} S^\nu +
                           \ldots \right] N + {\rm h.c.}
\end{eqnarray}
with
\begin{eqnarray}
f_{\mu\nu}^\pm &=& u^\dagger F_{\mu\nu}^R u \pm u F_{\mu\nu}^L
                   u^\dagger \equiv \tau^i \; f_{\pm\mu\nu}^i\; , \nonumber \\
F_{\mu\nu}^X   &=& \partial_\mu F_{\nu}^X - \partial_\nu F_{\mu}^X - i
                   \left[ F_{\mu}^X , F_{\nu}^X \right] ; \;\; X = L
                   , R \; , \nonumber \\
F_{\mu}^R      &=& {\bf v_\mu} + {\bf a_\mu} \;, \hspace{5mm} F_{\mu}^L \; =
                   \; {\bf v_\mu} - {\bf a_\mu} \; , \nonumber \\
v_{\mu\nu}^{(s)}&=& \partial_\mu v_{\nu}^{(s)} - \partial_\nu
                    v_{\mu}^{(s)} \; .
\end{eqnarray}
Note that in ${\cal L}_{\pi N}^{(2)}$ only two low--energy constants
(LECs) appear, which we have expressed in terms of the isoscalar and
isovector anomalous magnetic moments, $\kappa_s$ and $\kappa_v$, of the
nucleon, respectively. The LEC $b_1$ is fixed from neutral pion
photoproduction at threshold, $b_1 = 12.0$. The ${\cal O}(\epsilon^2)$
Lagrangians given so far are sufficient to calculate the leading order
$\Delta$(1232) effects in RMC. However, as mentioned above, in
OMC the leading $\Delta$(1232) related effects only occur at 
N$^2$LO---specifically as a ${\cal O}(\epsilon^3)$ contribution to the 
form factors of the nucleon. In SSE these terms have already been calculated 
in \cite{BFHM} and we refer the reader interested in the details of the
third order Lagrangian required for such a calculation to ref.\cite{BFHM}. 
We now have specified the effective Lagrangian needed to
work out OMC and RMC to second order in the small scale expansion.
The pertinent Feynman rules to perform these calculations are
collected in appendix~A.

%%%%%%%%%%%%%%%%% mu capture %%%%%%%%%%%%%%%%%%%%%%%%%%%%%%%%%%%%%%%%%%%
\section{Muon capture}
\label{sec:muon}
\def\theequation{\arabic{section}.\arabic{equation}}
\setcounter{equation}{0}

This section is concerned with the theoretical description of OMC and
RMC. To keep the manuscript self--contained, we give all formulae
necessary to calculate the capture rates.

\subsection{Leptonic matrix elements}

Consider first the purely leptonic part of the muon capture reaction.
For OMC and RMC, we need the following two leptonic matrix elements:
\begin{eqnarray}
\langle
\nu_\mu|J_\mu^+|\mu\rangle&=&-i\,\frac{g_2}{\sqrt{8}}\,\bar{\nu}_\mu(l^\prime)
                \gamma_\mu\left(1-\gamma_5\right)\mu(l)~, \\
\langle
\nu_\mu\,\gamma|J_\mu^+|\mu\rangle&=&-i\,\frac{g_2}{\sqrt{8}}\,\frac{e}{2
                l\cdot
k}\,\bar{\nu}_\mu(l^\prime)\gamma_\mu\left(1-\gamma_5\right)
                \left(2 \epsilon^\ast \cdot
l-\not{k}\not{\epsilon}^\ast\right)\mu(l)~,
\end{eqnarray}
with $\epsilon^\ast_\mu$ the polarization vector of the photon.
The corresponding four--momenta have already been defined in
Eq.(\ref{OMCdef}) and Eq.(\ref{RMCdef}), respectively.

\subsection{Hadronic matrix elements}
\label{sec:res}
\def\theequation{\arabic{section}.\arabic{equation}}

\begin{figure}[bht]
\centerline{
\epsfysize=1.5in %4.5
\epsffile{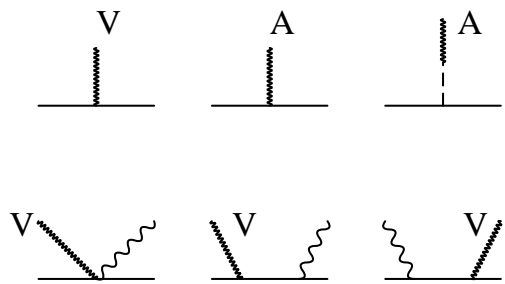}
}
\vskip 1cm %%3

\caption[]{Graphical representation of the V, A (first row),
and VV (second row) correlators.
Nucleons, pions, photons and external gauge fields are denoted by
solid, dashed, wiggly and zigzag lines, in order.}\label{fig:mu12}
\end{figure}

\begin{figure}[htb]
\centerline{
\epsfysize=3.5in %4.5
\epsffile{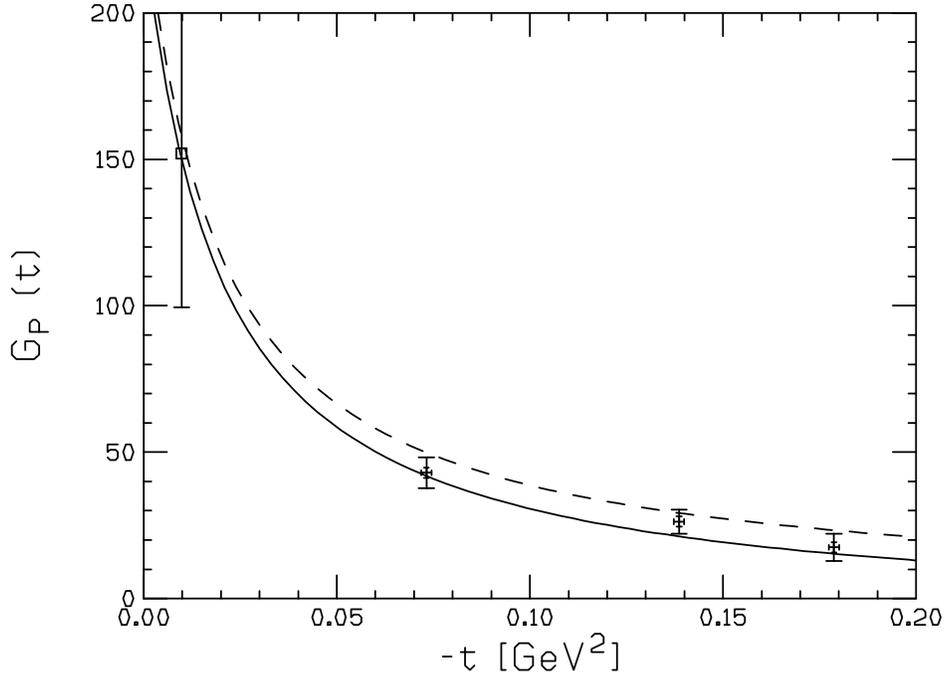}
}

\vskip 1cm %%3

\caption[]{The ``world data'' for the induced pseudoscalar form factor
$G_P(t)$.  Dashed curve: Pion--pole prediction. Solid curve:
${\cal O}(\epsilon^3)/{\cal O}(p^3)$ SSE/HBChPT prediction. 
The pion electroproduction data
(crosses) are from ref.\cite{choi}. Also shown is the OMC result
at $t = -0.88m_\mu^2$ from ref.\cite{bald} (open square).
}\label{fig:gp}
\end{figure}

For OMC one only needs to calculate the vector and axial-vector transition
matrix elements $\langle n|V_\mu^-|p\rangle,\;\langle n|A_\mu^-|p\rangle$
of Eqs.(\ref{strongME1},\ref{strongME2}). In the limit of exact isospin
symmetry these matrix-elements constitute the off-diagonal components
of the isovector nucleon vector and axial-vector currents which are typically
parameterized in terms of four form factors (e.g. ref.\cite{BFHM}): 
the isovector vector form factors $F_{1,2}^v$, as well as the axial and 
the induced
pseudoscalar form factor,  $G_A (q^2)$ and $G_P (q^2)$ respectively. 
In terms of these, the relativistic strong matrix elements
defined in
Eqs.(\ref{strongME1},\ref{strongME2}) 
(which corresponds to the vector and axial correlators, shown in  the first 
row of fig.\ref{fig:mu12}) are given by:
\begin{eqnarray}
\langle n|V_\mu^-|p\rangle&=&-i\,
                             \frac{g_2 V_{ud}}{\sqrt 8}\,\bar{n}(p_2)
                             \left[F_1^v(q^2)
\gamma_\mu +\frac {i}{2M_N}F_2^v(q^2) \sigma_{\mu \nu} q^\nu
\right]p(p_1)~, \label{Vcorr}  \\
\langle n|A_\mu^-|p\rangle&=&-i\,
                             \frac{g_2 V_{ud}}{\sqrt 8}\,\bar{n}(p_2)
                             \left[G_A (q^2)
\gamma_\mu\gamma_5+\frac{G_P(q^2)}{2M_N}q_\mu \gamma_5
\right]p(p_1). \label{Acorr} 
\end{eqnarray}
where $q^2=(p_2-p_1)^2$ is the invariant momentum transfer squared. The 
Dirac $F_1^{(v)}$  and Pauli $F_2^{(v)}$ form factor are subject to the 
normalizations
\begin{eqnarray}
F_1^{(v)}(0) =1 ~, \quad F_2^{(v)}(0) = \kappa_v~,
\end{eqnarray}
with $\kappa_v =3.71$ the isovector nucleon anomalous magnetic moment.
In the axial matrix element, Eq.(\ref{Acorr})  one has assumed the absence
of second class currents.
The electromagnetic form factor are rather well known. The current
situation concerning their theoretical understanding and experimental 
knowledge can be found e.g. in \cite{kmd}.
Muon captures therefore provide us with
the opportunity to study the weak axial structure of a nucleon.  
While $G_A (q^2)$ can be
extracted from (anti)neutrino--proton scattering or charged pion
electroproduction data, $G_P(q^2)$ is harder to pin down and in fact
{\em constitutes the least known nucleon form factor}.
In Fig.\ref{fig:gp} we present the  ``world data'' for $G_P(q^2)$.

These four form factors have already been calculated to ${\cal O}(\epsilon^3)$
in \cite{BFHM}. One finds\footnote{Note that due to the small momentum
transfer $|q^2|<0.01$ GeV$^2$ in OMC/RMC it is sufficient to work in the 
radius approximation of the form factors, i.e. to truncate the $q^2$-dependence
after the first term. The full ${\cal O}(\epsilon^3)$ momentum dependence 
for $-q^2<0.2$ GeV$^2$ can be found in \cite{BFHM}.}
\begin{eqnarray}
F_i^{(v)}(q^2)&=&F_i^{(v)}(0)\left[1+\frac{1}{6}(r_i^v)^2 q^2 + {\cal O}(q^4)
\right]  \label{ff1} \\
G_A(q^2)&=&g_A\left[1+\frac{1}{6}(r_A)^2 q^2 + {\cal O}(q^4) \right] 
           \label{ff2} \\
G_P(q^2)&=& \frac {4 M_N g_{\pi NN} F_\pi}{m_\pi^2 -q^2 }-\frac{2}{3}
g_A M_N^2 r_A^2\; . \label{gpff}
\end{eqnarray}
where we have systematically shifted all appearing quantities to their physical
values. $r_{1,2}^v$ and $r_A$ are the isovector Dirac and Pauli radius and
the axial radius respectively. They receive contributions from chiral loops
and counter-terms, except for $r_2^v$ which is free of any LEC. 
Detailed results can be found in \cite {BFHM}. 
In Fig.\ref{fig:gp} the difference between the usual pion-pole 
parameterization for $G_P(q^2)$ and the full chiral structure of the form
factor, Eq.\ref{gpff} is displayed. Note that this chiral structure
is not affected by $\Delta$(1232). 
In the kinematical region of RMC, which mainly lies to the
``left'' of the OMC point in Fig.\ref{fig:gp} the structure effect 
in $G_P(q^2)$ proportional to the axial radius $r_A$ is expected
to play only a small role. Certainly, the present experimental uncertainties
both in OMC~\cite{bald}
and in RMC~\cite{TRIUMF} are too large to distinguish between the two curves,
but new efforts are under way~\cite{proposal}\footnote{
Let us briefly emphasize that there exists another window on
$G_P(q^2)$---pion electroproduction. So far there has only been one
experiment~\cite{choi} that
took up the challenge, with the results shown in Fig.\ref{fig:gp}.
In this kinematical regime
the structure proportional to $r_A$ produces the biggest effect and a new
dedicated
experiment should be able to identify it --- thereby enhancing
our knowledge of this poorly known form factor considerably! In fact,
at the Mainz Microtron MAMI-B a dedicated experiment has been
proposed to measure the axial and the induced pseudoscalar form
factors by means of charged pion electroproduction at low momentum
transfer~\cite{MAMI}.}.

Usually the non relativistic reduction of Eqs.(\ref{Vcorr},\ref{Acorr})
is done in
the Breit frame. Here we give the non relativistic 
strong matrix elements in the rest frame of the proton where our
calculation is done:
\begin{eqnarray}
\langle n|V_\mu^-|p\rangle&=&-i\,\frac{g_2 V_{ud}}{\sqrt 8}\ {\cal N}_2 \,
\bar{n}_v(p_2) \left\{
\left(\frac{2 M_N}{E_2+M_N}F_1^{(v)}(q^2)
      -\frac{E_2-M_N}{E_2+M_N}F_2^{(v)}(q^2)\right)v_\mu \right.\nonumber\\
& &+\left[\frac{1}{E_2+M_N}\left(F_1^{(v)}(q^2)+F_2^{(v)}(q^2)\right)
      -\frac{1}{2 M_N}F_2^{(v)}(q^2)\right]q_\mu\nonumber\\
& &\left.+\frac{2}{E_2+M_N}\left[S_\mu,S\cdot q\right]
      \left(F_1^{(v)}(q^2)+F_2^{(v)}(q^2)\right)
\right\} p_v(0)~, \\
\langle n|A_\mu^-|p\rangle&=&-i\,\frac{g_2 V_{ud}}{\sqrt 8}\ {\cal N}_2 \,
\bar{n}_v(p_2)\left\{G_A (q^2)\left[2\,S_\mu - \frac{2\,S\cdot q\, v_\mu}
                  {E_2+M_N}\right] \right. \nonumber\\
& & \left. + R\,G_P(q^2)\,\frac{S\cdot q\,q_\mu}{M_N\left(E_2+M_N\right)}
\right\}p_v(0), 
\end{eqnarray}
where ${\cal N}_2$ is the usual normalization factor of the neutron wave
function,  ${\cal N}_2 =\sqrt{ \frac{E_2+M_N}{2M_N}}$ and $E_2$ is the
neutron energy.
One gets immediately the results to ${\cal O}(\epsilon^2)$ by replacing
the form factors by their values at $q^2=0$

\begin{figure}[bht]
\centerline{
\epsfysize=2.5in %4.5
\epsffile{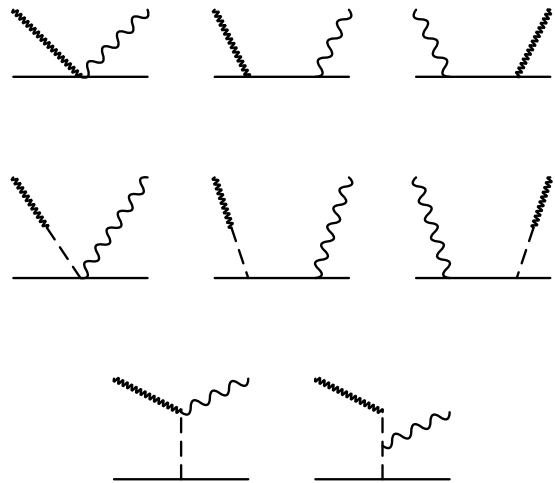}
}
\vskip 1cm %%3

\caption[]{Nucleon contributions to the VA correlator. All external
  gauge fields are axials.}\label{fig:mu3}
\end{figure}

\begin{figure}[htb]
\centerline{
\epsfysize=1.6in %4.5
\epsffile{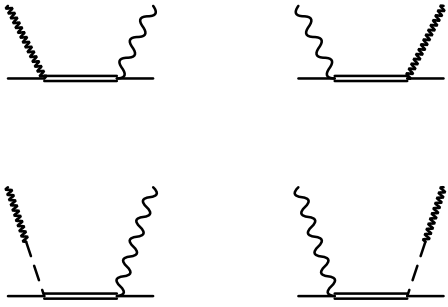}
}
\vskip 1cm %%3

\caption[]{Delta contributions to the VA correlator. All external
  gauge fields are axials.}\label{fig:mu4}
\end{figure}

We now turn to the vector--vector (VV) and vector--axial (VA) correlator
which we only need to ${\cal O}(\epsilon^2)$ in SSE.
Working in the  Coulomb gauge $\epsilon^\ast\cdot v= 0$ for the photon
and making use of the transversality condition $\epsilon^\ast\cdot
k=0$, we find (the pertinent
Feynman diagrams for the VV correlator are shown in the second row in
fig.~\ref{fig:mu12} and the ones for VA in figs.~\ref{fig:mu3},\ref{fig:mu4})
\begin{eqnarray}
\langle n|{\cal T} \,V\cdot\epsilon^\ast V_\mu^- |p\rangle^{(2)}&=&-i\,
                             \frac{g_2 V_{ud}\,
                             e}{\sqrt{8}}\,\bar{n}_v(r^\prime)\left\{
                             \frac{1+\kappa_v}{M_N}\left[S_\mu,S
                             \cdot\epsilon^\ast
                             \right]-\frac{1}{2 M_N}\,\epsilon^\ast_\mu
                             \right. \nonumber \\
& &\phantom{-i\,\frac{g_2 V_{ud}\,e}{\sqrt{8}}\,\bar{n}(r^\prime)} \left.
   +\frac{1}{M_N \omega}\,v_\mu \left[
   \left(1+\kappa_v\right)[S\cdot\epsilon^\ast,S\cdot k] -
   \epsilon^\ast\cdot r\right]+{\cal O}(1/M_N^2)
   \right\} p_v(r)~, \\
\langle n|{\cal T} \,V\cdot\epsilon^\ast A_\mu^- |p\rangle^{(2)}&=&-i\,
                             \frac{g_2 V_{ud}\,
                             e}{\sqrt{8}}\,\bar{n}_v(r^\prime) \times \nonumber
\\
& &\left\{2\,R\,g_A\,\frac{S\cdot(r^\prime-r)}{
   (r^\prime-r)^2-m_{\pi}^2}\times\left[\frac{2\,\epsilon^\ast\cdot(l-l^
   \prime)\,
   (l-l^\prime)_\mu}
   {(l-l^\prime)^2-m_{\pi}^2}-\epsilon^\ast_\mu\right]\right. \nonumber \\
& &- R\,\frac{g_A}{M_N}\,\frac{\left(v\cdot r^\prime-v\cdot r\right)
   S\cdot(r+r^\prime)}{(r^\prime-r)^2-m_{\pi}^2}
   \times\left[\frac{2\,\epsilon^\ast\cdot(l-l^\prime)\,(l-l^\prime)_\mu}
   {(l-l^\prime)^2-m_{\pi}^2}-\epsilon^\ast_\mu\right]
   \nonumber \\
& &-2\,R\,g_A\left[1+\frac{v\cdot l-v\cdot l^\prime}{2 M_N}\right]\frac{S\cdot
   \epsilon^\ast\,(l-l^\prime)_\mu}{(l-l^\prime)^2-m_{\pi}^2} +
\frac{g_A}{M_N}\,S\cdot
   \epsilon^\ast\,v_\mu \nonumber \\
& &+\frac{g_A}{M_N}\left[\frac{(2+\kappa_s+\kappa_v)\,
   S^\alpha\,[S\cdot\epsilon^\ast,S\cdot
k]}{\omega}+\frac{(\kappa_v-\kappa_s)\,
   [S\cdot\epsilon^\ast,S\cdot k]\,S^\alpha}{\omega}\right.\nonumber \\
& &\phantom{+\frac{g_A}{M_N}} \left.
   -\frac{2\,S^\alpha\epsilon^\ast\cdot r}{\omega}\right]
   \times\left[g_{\mu\alpha}-R\,\frac{(l-l^\prime)_\alpha(l-
   l^\prime)_\mu}{(l-l^\prime)^2-m_\pi^2}\right] \nonumber \\
& &+\frac{g_{\pi N\Delta}b_1}{3\, M_N}\left[\frac{2\Delta\,[k^\alpha
S\cdot\epsilon^\ast
   -\omega\, v^\alpha S\cdot\epsilon^\ast-\epsilon^{\ast\,\alpha}S\cdot
k]}{\Delta^2-
   \omega^2}-\frac{4\,S^\alpha[S\cdot\epsilon^\ast,S\cdot
k]}{3\,(\Delta+\omega)}
   \right. \nonumber \\
& &\phantom{+\frac{g_{\pi N\Delta}b_1}{3\, M_N}} \left. \left.
    +\frac{4\,[S\cdot\epsilon^\ast,S\cdot
k]\,S^\alpha}{3\,(\Delta-\omega)}\right]
   \times\left[g_{\mu\alpha}-\frac{(l-l^\prime)_\alpha(l-l^\prime)_\mu}{(l-l^
   \prime)^2-
   m_\pi^2}\right] +{\cal O}(1/M_N^2)\right\} p_v(r)~,\label{VAcorr}
\end{eqnarray}
with $\omega = v \cdot k$ and $R=1$ in QCD. We have introduced this factor
multiplying the Born term contributions proportional to the induced
pseudoscalar
form factor for the later discussion. One can easily check from the 
continuity equations satisfied by the correlators 
(which for example relates the vector--axial correlator to the 
axial one, see \cite{STK})
that gauge invariance is satisfied in the above equations.
We note again that here we only give the results to
${\cal O}(\epsilon^2)$ as this is sufficient to study the leading order 
$\Delta$(1232)
effects in RMC. We also want to point out
that the vector--vector correlator is free of delta
effects to this order ${\cal O}(\epsilon^2)$, i.e. the leading $\Delta$(1232)
effect only appears in the
vector--axial correlator, cf. fig~\ref{fig:mu4}.

%%%%%%%%%%%%%%%%%%%%% OMC %%%%%%%%%%%%%%%%%%%%%%%%%%%%%%%%%%%%%%%%%%%%%%%%%%%
\subsection{Ordinary Muon Capture}
\label{sec:omc}

Surprisingly, no full analysis of OMC exists in the literature for the
framework of
chiral effective field theories. All previous analyses \cite{bkmgp,BFHM,FLMS}
stopped
at the level of writing down the relevant matrix-elements/form factors
(here Eqs.(\ref{Vcorr},\ref{Acorr})), but no discussion of the
implications for the lifetime
of OMC was given, which can be written down in a closed analytic form.
In this section we are going to fill this gap.
We work in the Fermi approximation of a static $W_\mu^-$ field,
i.e. the gauge boson propagator is reduced to a point interaction
(since the typical momenta involved are much smaller than the $W$ mass),
\begin{eqnarray}
{\cal M}_{\mu^-p\rightarrow \nu_\mu n}&=&{\cal M}^{\rm OMC} =
                \langle \nu_\mu|W_\mu^+|\mu\rangle
                \,i\,\frac{g^{\mu\nu}}{M_W^2}\left[\langle n|V_\nu^-|p\rangle -
                \langle n|A_\nu^-|p\rangle\right]~.
\end{eqnarray}

\subsubsection{Spin-averaged OMC}

Introducing the Fermi constant $G_F$ via $G_F={g_2^2\sqrt{2}}/{(8 M_W^2)}$,
we define the square of the {\em spin-averaged} invariant matrix element to be
\begin{eqnarray}
\frac{1}{4}\sum_{\sigma\sigma^\prime s s^\prime}|{\cal M}^{\rm OMC}|^2&=&
\frac{G_F^2 V_{ud}^2}{2}\, L_{\mu\nu}^{(a)}H^{\mu\nu}_{(a)} \; .
\end{eqnarray}
With the normalizations
\begin{eqnarray}
\sum_s \mu(l,s) \bar{\mu}(l,s)=\frac{\not{l}+m_{\mu^-}}{2\, m_{\mu^-}} ,&&
          \quad  \sum_s \nu(l^\prime,s) \bar{\nu}(l^\prime,s)=\not{l}^\prime ,
\nonumber \\
\sum_\sigma p_v(r,\sigma) \bar{p}_v(r,\sigma)=P_v^+\left(1+\frac{v\cdot r}{2
M_N}\right)
            ,\quad && \sum_\sigma n_v(r^\prime,\sigma)
\bar{n}_v(r^\prime,\sigma)=P_v^+
            \left(1+\frac{v\cdot r^\prime}{2 M_N}\right) ,
\end{eqnarray}
one then obtains the symmetric tensors
\begin{eqnarray}
L_{\mu\nu}^{(a)}&=&\frac{2}{m_{\mu^-}}
                   \left\{l_\mu l^\prime_\nu-g_{\mu\nu}l\cdot
l^\prime+l^\prime_\mu
                   l_\nu+i\epsilon_{\mu\alpha\nu\beta}l^\alpha l^{\prime\beta}
                   \right\}~, \\
%H_{\mu\nu}^{(a)}&=&v_\mu v_\nu +g_A^2\left(v_\mu v_\nu-g_{\mu\nu}\right)
%                   -g_A^2\frac{\vec{r}^{\,\prime\, 2}+2 m_\pi^2}
%                   {(\vec{r}^{\,\prime\, 2}+m_\pi^2)^2}\,r^\prime_\mu
%r^\prime_\nu
%                   \nonumber \\
%                & &+\frac{1+g_A^2}{2 M_N}\left(v_\mu r^\prime_\nu+r^\prime_\mu
%v_\nu
%                   \right)
%                   +\frac{g_A(1+\kappa_v)}{M_N}\,i\epsilon_{\mu\nu}^{\alpha\beta}
%
%                   r^\prime_\alpha v_\beta+{\cal O}(1/M_N^2)~.
H_{\mu\nu}^{(a)}&=&a v_\mu v_\nu +b\left(v_\mu v_\nu-g_{\mu\nu}\right)
                   +c  \,r^\prime_\mu r^\prime_\nu
          +   d \,\left(v_\mu r^\prime_\nu+r^\prime_\mu v_\nu  \right)
                   +e\,i\epsilon_{\mu\nu}^{\alpha\beta}
                   r^\prime_\alpha v_\beta ~.
\end{eqnarray}
Note that we have evaluated the tensors for the special kinematic condition of
both the
proton and the muon being at rest\footnote{This approximation is well-justified
due to
the low binding energy $E\approx 2.5\,$keV of an s-wave muonic atom as compared
to the
muon mass.}, i.e. $l_\mu=(m_{\mu^-},0,0,0),\; r_\mu=(0,0,0,0)$.
$a, b, c, d, e$ are coefficients which depend on the four form factors 
Eqs.(\ref {ff1}, \ref {ff2}). For illustration we give them 
to ${\cal O}(\epsilon^2)$ 
\begin{eqnarray}
a= 1\, ,  \quad 
b=g_A^2 \, , \quad 
c= -g_A^2\frac{\vec{r}^{\,\prime\, 2}+2 m_\pi^2}
                   {(\vec{r}^{\,\prime\, 2}+m_\pi^2)^2}\, , \quad 
d =\frac{1+g_A^2}{2 M_N} \, ,\quad 
e=\frac{g_A(1+\kappa_v)}{M_N}~. 
\end{eqnarray}

We now assume that the initial muon-proton system constitutes the ground-state
of a bound
system. We therefore replace the plane-wave wavefunction of the muon used so
far in the
calculation by the 1s Bohr-wavefunction $\Phi(x)_{1s}$ of a muonic atom.
We note
that to ${\cal O}(\epsilon^2)$ we are not sensitive to the extended
structure
of the proton -- any dependence of the capture process on the electric/magnetic
radius of
the nucleon will only occur at ${\cal O}(\epsilon^3)$ \cite{BFHM}. As we will
see below, the terms of ${\cal O}(\epsilon^3)$ in OMC only represent
a small N$^2$LO correction. For simplicity we therefore approximate
the muonic atom wave-function dependence by its value at the origin even
at ${\cal O}(\epsilon^3)$, 
which effectively constitutes an upper bound for the ${\cal O}(\epsilon^3)$ 
correction to the total width. In a systematic ${\cal O}(\epsilon^3)$ 
analysis one would of course have to calculate
the overlap between the Bohr-wavefunction and the respective proton-neutron
transition form factors. Confining ourselves to $x=0$ we use
\begin{equation}
\Phi(0)_{1s}= \frac{\alpha^{3/2}\mu^{3/2}}{\sqrt{\pi}}\; ,
\end{equation}
with the reduced mass $\mu={M_N m_{\mu^-}}/{(M_N+m_{\mu^-})}$ and
$\alpha={e^2}/{4\pi\hbar c}$ in the Heaviside-Lorentz convention.
We can therefore calculate the spin-averaged rate of ordinary muon capture via
\begin{eqnarray}
\Gamma_{\rm OMC}&=&|\Phi(0)_{1s}|^2 \int\frac{d^3r^\prime}
               {(2\pi)^3 J_n}\,\frac{d^3 l^\prime}{(2\pi)^3 J_\nu}
               \left(2\pi\right)^4\delta^4\left(r+l-r^\prime-l^\prime\right)
               \frac{1}{4}
               \sum_{\sigma\sigma^\prime s s^\prime}|{\cal M}^{\rm OMC}|^2
\end{eqnarray}
with the normalization factors $J_\nu=2 E_\nu,\,J_n=1+v\cdot r^\prime/M_N$.
Evaluating all expressions at ${\cal O}(\epsilon^3)$ accuracy
one obtains 
\begin{eqnarray}\label{OMCrate}
\Gamma_{\rm OMC}^{\rm SSE}&=&\left( \,247.0\,-\,61.6
\,-4.0+{\cal O}(1/M_N^3)\right)\times
                          {\rm s}^{-1}\nonumber \\
            &=&181.5\times {\rm s}^{-1} 
\end{eqnarray}
Here we have used the coupling constants, masses and radii as given in
table~\ref{tab:val}. $r_1^{(v)}$ and $r_A$ corresponds to their empirical 
values since the value of the counter-terms which enter in their expressions
to ${\cal O}(\epsilon^3)$
has been set to reproduce them. The value of 
$(r_2^{(v)})^2_{\rm SSE}=0.61$ fm$^2$ is given
counter-term free by the ${\cal O}(\epsilon^3)$ SSE calculation. 
For comparison we also give the spin averaged OMC capture rate to 
${\cal O}(p^3)$ in HBCHPT without explicit spin 3/2 degrees of freedom:
\begin{eqnarray}\label{OMCratep3}
\Gamma_{\rm OMC}^{HBChPT}&=&\left( \,247.0\,-\,61.6
\,-3.8+{\cal O}(1/M_N^3)\right)\times
                          {\rm s}^{-1}\nonumber \\
            &=&181.7\times {\rm s}^{-1} 
\end{eqnarray}
The only change occurs at N$^2$LO via $(r_2^{(v)})^2_{\rm HBChPT}=0.52$ fm$^2$,
showing only a very weak dependence on the exact value of the isovector
Pauli radius, which has the ``exact'' value $(r_2^{(v)})^2_{exp.} =0.80$ 
fm$^2$.
We note that in Eqs.(\ref{OMCrate})
the ${\cal O}(\epsilon^2)$ SSE  
contribution amounts to a correction of 25\% of
the leading term, while the ${\cal O}(\epsilon^3)$ one is two 
orders of magnitude smaller,
indicating that muon capture on a proton really constitutes a
system with an extremely well behaved chiral expansion.

Finally we discuss the case of no explicit chiral symmetry breaking
(i.e. in the chiral limit $m_\pi = 0$). One expects
the spin-averaged capture rate to behave as
\begin{eqnarray}
\Gamma_{\rm OMC}^{\chi}&=&\left(\,214\,-\,46\,+{\cal O}(1/M_N^2)\right)
                          \times {\rm s}^{-1}\nonumber \\
                       &=&168 \times {\rm s}^{-1} \; .
\end{eqnarray}
which only represents a small finite shift compared to the value for finite
quark masses. The leading infrared singularities only occur at N$^2$LO
due to the long range nature of the chiral pion cloud in the isovector 
form factor radii \cite{BFHM}.

The reason for the nice stability of perturbative calculations for 
OMC in the physical world of small finite quark masses is of course
the fact that contributions of order $n$ are suppressed by
$\left(m_i/\Lambda_\chi\right)^{n-1}$, with $i=\pi,\mu$ and
$\Lambda_\chi\sim M_N\sim 1$~GeV. Note that one could also retain the
higher order {\it kinematical} corrections starting at order $1/M_N^3$.
To be more precise, this refers to the energy--momentum relation between the
various particles, the mass term appearing in the various projection
operators and the phase space. In that case, only the various correlators
(A,V,AV,VV) are truncated at order $1/M_N$. Using this approximation,
the total rate is  $\Gamma_{\rm OMC} = 193\,{\rm s}^{-1}$,
not much different from the one given in Eq.(\ref{OMCrate}).
We remark already at this point that in the case of RMC, it is mandatory
to retain these higher order terms if one stays at a low order in the
small scale expansion as done here.

\subsubsection{Hyperfine effect in OMC}

The spin-averaged OMC scenario presented in the previous section is only of
theoretical
interest. In nature the weak-interactions shows a very strong spin-dependence,
which
leads to quite different decay-rates depending on whether the captured 1s muon
forms a
singlet or a triplet spin-state with its proton \cite{Opat}, where the singlet
is the
usual state $(1/\sqrt{2})(|\uparrow,\downarrow \rangle
- |\downarrow, \uparrow\rangle)$ in terms of the muon and proton spins and the
triplet accordingly. Although the hyperfine splitting between
the two levels ``only'' amounts to 0.04~eV, the occupation numbers of the levels
due to thermal and collision induced processes tend to be far from statistical
equilibrium. In order to make any contact with experiment, 
the
singlet/triplet rates need to be calculated
separately. These various spin states are obtained
by using the following projection operators for the muon,
\begin{eqnarray}
\mu(l,1/2) \bar\mu(l,1/2) &=& \frac{1}{2}\bigl( 1 + \gamma_5 \barr s \bigr)\,
\frac{\barr{l}+m_\mu}{2m_\mu}~,\nonumber \\
\mu(l,\pm 1/2) \bar\mu(l, \mp 1/2) &=& \frac{\barr{l}+m_\mu}{2m_\mu (E_\mu +
m_\mu)}
\, \gamma_5  \gamma_0  \frac{1}{2} (\gamma_1 \pm i\gamma_2)   \gamma_0 \,
(\barr{l}+m_\mu)~,
\end{eqnarray}
and similarly for the proton,
\begin{eqnarray}
u(l,1/2) \bar u(l,1/2) &=& \frac{1}{2}\bigl( 1 + \gamma_5 \barr s \bigr)\,
\frac{1}{2} (1+\barr v)\biggl( 1 + \frac{v\cdot r}{2M_N}\biggr)~,\nonumber \\
u(l,\pm 1/2) \bar u(l, \mp 1/2) &=& \frac{1}{2}(1+\barr v)\,
\frac{\barr{r}+2M_N}{2M_N (E_N + M_N)}
\, \gamma_5  \gamma_0  \frac{1}{2} (\gamma_1 \pm i\gamma_2)   \gamma_0 \,
(\barr{r}+2M_N)\, \frac{1}{2}(1+\barr v)~.
\end{eqnarray}
For the total capture rates of singlet and the triplet states in the muonic
atom, we then find the following
decomposition into leading, next--to--leading order pieces and
 next--to--next--to--leading order pieces
to ${\cal O}(\epsilon^3)$ in SSE\footnote{For the remaining part of the 
discussion on OMC we only give the values calculated in to 
${\cal O}(\epsilon^3)$ in the small scale expansion. The corresponding
values to ${\cal O}(p^3)$ in HBChPT are very similar. For OMC the two
effective field theories only differ in the N$^2$LO contribution of the
isovector Pauli radius as discussed in the previous section.}:  
\begin{eqnarray}
\Gamma^{\rm sing}_{\rm OMC} &=& (957 - 245\,{\rm GeV}/M_N+ (30.4\,{\rm GeV}/M_N^2
-43.17)+{\cal O}(1/M_N^3))\times {\rm
  s}^{-1}=687.4\times {\rm  s}^{-1}~,\nonumber\\
\Gamma^{\rm trip}_{\rm OMC} &=& (10.3 + 4.72\,{\rm GeV}/M_N - 
(1.22\,{\rm GeV}/M_N^2 +1.00)+{\cal O}(1/M_N^3))
\times {\rm s}^{-1}=12.9\times {\rm s}^{-1}~,\label{Vero}
\end{eqnarray}
displaying the {\em dramatic spin-dependence} due to the V-A structure of the
weak
interaction in the Standard Model. These numbers correspond to a value of
$g_{\pi NN}$ as given in table~\ref{tab:val} and includes the pion pole 
corrections, Eq.(\ref{gpff}).
Since $G_P$ contributes negatively to the singlet rate a larger
value of $g_{\pi NN}$ leads to a smaller value for the rate: $\Gamma^{\rm sing}_{\rm OMC}=681.9\times {\rm  s}^{-1}$ for $g_{\pi NN}=13.4$. Similarly,
neglecting the pion pole corrections leads to 
$\Gamma^{\rm sing}_{\rm OMC}=676.1\times {\rm  s}^{-1}$.
In Eq.(\ref{Vero}) we have split 
the ${\cal O}(\epsilon^3)$  term (third and fourth terms
in parenthesis) into the contribution from the $1/M^2$ corrections and the
pure ${\cal O}(\epsilon^3)$ terms stemming from the various radii
which lead to the $q^2$ dependence of the
 form factors, Eq.(\ref{gpff}). It is very interesting
to note that these two contributions more or less cancels themselves in 
the case of the singlet term. It is thus extremely important to 
perform a consistent chiral expansion. In a relativistic Born model one obtains
the simple nice formula\footnote{We note that we have different relative 
signs compared to ref.\cite{san} to translate into our notation.} 
for $\Gamma^{\rm sing}_{\rm OMC}$ \cite{san}:
\begin{eqnarray}
\Gamma^{\rm sing}_{\rm OMC}&\sim&\left(6.236\,F_1^v(q_0^2)+0.5513\,F_2^v(q_0^2)
+16.44\, G_A(q_0^2) -0.2834\,G_P(q_0^2) \right)^2   \label{san} \\
&\sim&683\times {\rm s}^{-1}
\end{eqnarray}
where $q_0^2=-0.88 m_\mu^2$ is the momentum transfer and the remaining 
parameters are taken from table~\ref{tab:val}. The smaller value quoted
in \cite{prim}, \cite{san} comes from the fact that in the sixties $g_A$
was somewhat smaller and $r_A$ somewhat bigger.
This formula though
very appealing should not be used, being in contradiction with the
modern viewpoint of power counting. The good agreement with the SSE result
is purely accidental.
The problem arises from the fact that 
$\Gamma^{\rm sing}_{\rm OMC}$  is a rather sensitive quantity as can be seen
for example in Eq.(\ref{san}). Indeed the terms proportional to  
$F_2^v(q_0^2)$ and $G_P(q_0^2)$ are of the same order of magnitude
but have different signs so they have a tendency to cancel each other
rendering the values of $\Gamma^{\rm sing}$ rather sensitive to the
exact values of these two quantities. 

So far we have only considered OMC for the case of muonic atoms. For the
case of a
liquid hydrogen target one also has to take into account the possibility of
muon
capture in a muonic molecule $p\mu p$, which can be formed via the reaction
$p\mu+pep\rightarrow p\mu p+e+124$eV. In such a molecule the muon can be found
in
a so called ortho $(O)$ (spin of the protons parallel) or para $(P)$
(spin of the protons antiparallel) spin state relative to its two accompanying
protons.
The decay
rates of these molecular states can be calculated from the singlet/triplet
rates of the
muonic atom via
\begin{eqnarray}
\Gamma_P &=&2\gamma_P \frac{1}{4}(3\Gamma_{\rm trip} + \Gamma_{\rm sing} )~,
\label{para} \\
\Gamma_O &=&p_{1/2}\,\Gamma_{1/2}+p_{3/2}\,\Gamma_{3/2}~, \label{ortho}
\end{eqnarray}
The wavefunction corrections\footnote{$\gamma_P\;(\gamma_O)$ denotes the ratio
of
the probability of finding the
negative muon at the point occupied by a proton in the para-muonic
(ortho-muonic)
molecule and the probability of finding the negative muon at the origin in the
muonic atom. We are grateful to Shung-Ichi Ando for pointing out ref.\cite{WP}
to us.}
are taken to be $\gamma_O=0.500,\;\gamma_P=0.5733$ \cite{WP}.
We note that the {\em para molecular state} is often referred to as the
statistical
mixture,
as it corresponds to the naively expected occupation numbers of the muonic
atom. For a
precise calculation of the {\em ortho molecular state} on the other hand one
first
has to know the exact probabilities $p_{1/2},p_{3/2}$ for the muonic molecule
being in a total spin S=(1-1/2)=1/2 or
a total spin S=(1+1/2)=3/2 state, with $p_{1/2}>0.5$ \cite{Weinberg}.
The corresponding decay rates are given by
\cite{prim,Weinberg}
\begin{eqnarray}
\Gamma_{1/2}&=&2\gamma_O\left(\frac{3}{4}\Gamma_{\rm sing}+\frac{1}{4}
\Gamma_{\rm trip}\right)\nonumber\\
\Gamma_{3/2}&=&2\gamma_O\;\Gamma_{\rm trip}
\end{eqnarray}
%Given that $p_{1/2}>0.5$ \cite{Weinberg}
%and $\Gamma_{\rm sing}\gg\Gamma_{\rm trip}$ for OMC we arrive at the formulae
%\begin{eqnarray}
%\Gamma_P^{\rm OMC} &=&2\gamma_P \frac{1}{4}(3\Gamma_{\rm trip} + \Gamma_{\rm sing} )~,  \\
%         &=&(283-66.3\,{\rm GeV}/M_N-2.61\,{\rm GeV}^2/M_N^2 
%            +{\cal O}(1/M_N^3)) \times {\rm s}^{-1}=210
%            \times {\rm s}^{-1}\nonumber\\
%\Gamma_O^{\rm OMC} &\approx&2\gamma_O\;\frac{3}{4}\Gamma_{\rm sing}\nonumber\\
%         &\approx& (718 - 184\,{\rm GeV}/M_N-2.17\,{\rm GeV}^2/M_N^2
%          +{\cal O}(1/M_N^3)) \times {\rm s}^{-1}=
%               520    \times {\rm s}^{-1}\; .\label{orthosimple}
%\end{eqnarray}
Theoretical calculations of the spin-effects in the muonic molecule
\cite{BV} suggest $p_{1/2}\approx 1,\;p_{3/2}\approx 0$
which leads to the values:
\begin{eqnarray}
\Gamma_P^{\rm OMC} &=& 208\times {\rm s}^{-1}\nonumber\\
\Gamma_O^{\rm OMC} &=& 493 \cdots 519   \times {\rm s}^{-1}\; .\label{orthosimple}
\end{eqnarray}
where the range given in $\Gamma_O^{\rm OMC}$ corresponds to $0.95 \leq p_{1/2}
\leq 1$. We have allowed here for a 5\% uncertainty in the occupation
numbers\footnote{The possibility  of very different
occupation numbers has been raised by Shung-Ichi Ando during the
Chiral Dynamics 2000 conference. This will be reported in a forthcoming
paper, see ref.\cite{AKM}.}  
 to show the sensitivity of our results on this quantity. Since
$\Gamma_{\rm sing}\gg\Gamma_{\rm trip}$ for OMC, $\Gamma_O^{\rm OMC}$ turns out to be
roughly proportional to $ p_{1/2}$.
We note that our number for
capture from the molecular ortho state agrees very well with the most recent
measurement $\Gamma_O^{\rm exp.}=(531\pm33)\times {\rm s}^{-1}$ from Bardin et al.
\cite{bald}. Let us stress at this point the importance of the new 
proposed experiment at PSI \cite{proposal} which will be done with a
hydrogen gas target and will thus be independant on these occupation numbers.
One will directly measure $\Gamma_{sing}$.

With these expressions given above one can now calculate the rate for muon
capture in
liquid hydrogen for a general scenario, if one knows the relative occupation
numbers
for the atomic singlet\footnote{We
assume that all muonic atoms initially in the triplet state are effectively
converted
into the atomic singlet
state through collision with hydrogen molecules in the stopping target via the
Gershtein-Zeldovich mechanism
\cite{prim}.} $f_S$, the
molecular ortho $f_O$ and the molecular para $f_P$ state:
\begin{equation}
\Gamma_{\rm OMC}^{H_2}=f_S\;\Gamma_{\rm sing}+f_O\;\Gamma_O+f_P\;\Gamma_P
\end{equation}
For example, in the case of the recent TRIUMF experiment with
$f_S=0.061,\,f_O=0.854,\,
f_P=0.085$ \cite{TRIUMF} one would obtain a total capture rate of
$\Gamma_{\rm OMC}^{\rm TRIUMF}=(504+{\cal O}(1/M_N^3))\times {\rm s}^{-1}$.

%%%%%%%%%%% RMC %%%%%%%%%%%%%%%%%%%%%%%%%%%%%%%%%%%%%%%%%%%%%%%%%%%%%%%%
\subsection{Radiative muon capture}\label{sec:four}
\subsubsection{Total capture rates}
In the static approximation for the W--boson, the pertinent matrix element
for RMC decomposes into two terms,
\begin{eqnarray}
{\cal M}_{\mu^-p\rightarrow \nu_\mu n\gamma}&=&\langle
\nu_\mu|W_\mu^+|\mu\rangle
                \,i\,\frac{g^{\mu\nu}}{M_W^2}\left[
                \langle n|{\cal T} \,V\cdot\epsilon^\ast V_\nu^- |p\rangle
                -\langle n|{\cal T} \,V\cdot\epsilon^\ast A_\nu^- |p\rangle
                \right] \nonumber \\
             & &+\langle \nu_\mu\,\gamma|W_\mu^+|\mu\rangle
                \,i\,\frac{g^{\mu\nu}}{M_W^2}\left[\langle n|V_\nu^-|p\rangle -
                \langle n|A_\nu^-|p\rangle\right]~,
\end{eqnarray}
so that its square in the spin--averaged case can be written as a sum of
four terms, with both photons coming either from the hadronic or the
leptonic side and two mixed terms, i.e.
\begin{eqnarray}
\frac{1}{4}
\sum_{\sigma\sigma^\prime s s^\prime\lambda\lambda^\prime}|{\cal M}^{\rm RMC}|^2&=&
\frac{e^2 G_F^2 V_{ud}^2}{2}\left[L_{\mu\nu}^{(a)}H^{\mu\nu}_{(d)}
+\left(\sum_{\lambda\lambda^\prime}L_{\mu\nu}^{(b)}H^{\mu\nu}_{(c)}+
L_{\mu\nu}^{(c)}H^{\mu\nu}_{(b)}\right)
+L_{\mu\nu}^{(d)}H^{\mu\nu}_{(a)}\right]~,
\end{eqnarray}
with $\lambda , \lambda'$ the photon helicities.
Explicit expressions for the various tensors are not given here
because they are lengthy and not illuminating. We also note that
standard packages like REDUCE can not be used straightforwardly
to obtain these tensors since the cyclicity of the trace in the
presence  of $\gamma_5$ matrices is not fulfilled.
The total decay rate is given by:
\begin{equation}\label{RMCrate}
\Gamma_{\rm tot}
= \frac{ |\Phi(0)_{1s}|^2 }{16\pi^4} \,
 \int_0^\pi  \sin\theta d\theta \, \int_0^{\omega_{\rm max}}
 d\omega \, \omega \, l_0'\,
\biggl( 1 - \biggl( \frac{m_\mu - \omega (1 - \cos\theta)}{M_N}\biggr)\biggr)\,
\frac{1}{4}\, \sum_{\sigma\sigma^\prime s s^\prime\lambda\lambda^\prime}
|{\cal M}^{\rm RMC}|^2~,
\end{equation}
with $\omega = k_0$ the photon energy. The direction of the photon defines
the z--direction and $\theta$ in Eq.(\ref{RMCrate}) is the polar angle of
the outgoing lepton with respect to this direction. The maximal photon energy
is given by
\begin{equation}
\omega_{\rm max} = m_\mu\,\biggl( 1 + \frac{m_\mu}{2M_N}\biggr) \,
\biggl( 1 + \frac{m_\mu}{M_N}\biggr)^{-1} ~.
\end{equation}
Furthermore, the energy of the outgoing lepton follows from energy
conservation,
\begin{equation}\label{l0}
l_0'  = m_\mu  - \omega -\frac{m_\mu^2}{M_N} + \frac{\omega(1-\cos\theta)(m_\mu
- \omega)}{M_N} + {\cal O}(1/M_N^2)~.
\end{equation}
First we discuss the (academic) spin-averaged RMC scenario, which allows for
a comparison with previous calculations:
\begin{eqnarray}
\Gamma^{\rm RMC}_{\rm spin av.}&=&\left(66.0+18.7+{\cal O}(1/M_N^2)\right)
\times10^{-3}\,s^{-1}=84.7\times10^{-3}\,s^{-1}\;{\rm (HBChPT)}\nonumber\\
\Gamma^{\rm RMC}_{\rm spin av.}&=&\left(66.0+20.4+{\cal O}(1/M_N^2)\right)
\times10^{-3}\,s^{-1}=86.4\times10^{-3}\,s^{-1}\;{\rm (SSE)} .
\end{eqnarray}
Both the HBChPT and the SSE results suggest a good convergence for the chiral
expansion, as expected from dimensional analysis. Note that the leading order
capture
rates in both calculations are identical,
as $\Delta$(1232) related effects only start at sub-leading order. Our leading
order
result also agrees\footnote{Ref.\cite{meiss} gives
$\Gamma^{\rm RMC,\,HBChPT}_{\rm spin av.} = 61\times 10^{-3}\,{\rm s}^{-1}$ to leading
order.
The small difference can be traced back to the different values of some
of their input parameters.} with the HBChPT calculation of ref.\cite{meiss}. 

However, our HBChPT ${\cal O}(p^2)$ correction is nearly 30\% larger than the 
one given in \cite{meiss}.
Comparing with the corresponding ${\cal O}(\epsilon^2)$ correction in SSE, we
note that
$\Delta$(1232) indeed leads to a larger decay RMC decay rate and constitutes a
9\%
correction to our ${\cal O}(p^2)$ contribution. However, the total
(spin-averaged)
decay rate is only affected by 2\% {\em due to the fast convergence of the
chiral series} for RMC.

For the case of muonic atoms we obtain the following decay rates in the
singlet/triplet
channel\footnote{Note the reversal of
  the relative size of the singlet to triplet contribution as compared
  to the case of OMC.}, utilizing the projection formalism outlined in
sec.\ref{sec:omc}.
\begin{eqnarray}
\Gamma^{\rm RMC}_{\rm sing}&=&\left(12.7-18.7\,{\rm GeV}/M_N+{\cal O}(1/M_N^2)\right)
\times10^{-3}\,s^{-1}=3.10\times10^{-3}\,s^{-1}\;{\rm (HBChPT)}\nonumber\\
\Gamma^{\rm RMC}_{\rm sing}&=&\left(12.7-18.3\,{\rm GeV}/M_N+{\cal O}(1/M_N^2)\right)
\times10^{-3}\,s^{-1}=2.90\times10^{-3}\,s^{-1}\;{\rm (SSE)}\\
& &\nonumber \\
\Gamma^{\rm RMC}_{\rm trip}&=&\left(119-3.86\,{\rm GeV}/M_N+{\cal O}(1/M_N^2)\right)
\times10^{-3}\,s^{-1}=112\times10^{-3}\,s^{-1}\;{\rm (HBChPT)}\nonumber\\
\Gamma^{\rm RMC}_{\rm trip}&=&\left(119-1.80\,{\rm GeV}/M_N+{\cal O}(1/M_N^2)\right)
\times10^{-3}\,s^{-1}=114\times10^{-3}\,s^{-1}\;{\rm (SSE)}
\end{eqnarray}
Note that for the total numbers given we did
not expand all kinematical factors in powers of $1/M_N$ since in case
of the small singlet, the contribution from the terms starting at
order  $1/M_N^2$ can not be neglected. In fact, a strict truncation
at $1/M_N$ leads to an unphysical negative singlet capture rate.
For the much bigger triplet, these higher order corrections are much
less important. We remark that in  ref.\cite{meiss}  no strict
 $1/M_N$ expansion was performed, only at some places these authors
used the leading order results, e.g. for the nucleon energy by neglecting
the recoil term.
Only if one goes to a sufficiently high order in the small scale expansion,
the truncation of these kinematical factors can be justified. Comparing
the ${\cal O}(p^2)$ HBChPT with the ${\cal O}(\epsilon^2)$ SSE calculation,
we observe that while the total  
singlet capture rate is nearly identical in both approaches, the 
absolute value of the $1/M_N$ term in the total triplet capture rate is a
factor of two different between SSE and HBChPT. However this NLO term is much 
smaller than the LO triplet rate, leading to a rather similar total 
spin triplet RMC rate  with or
without explicit $\Delta$(1232) contributions.

Finally we address the complications for RMC due to the presence of muonic
molecules in the liquid hydrogen target. According to Eq.(\ref{para}), we can
easily
determine the capture rate from the molecular para state
\begin{eqnarray}
\Gamma^{\rm RMC}_{P}&=&85.2\times10^{-3}\,s^{-1}\;{\rm (HBChPT)} \nonumber \\
\Gamma^{\rm RMC}_{P}&=&86.4\times10^{-3}\,s^{-1}\;{\rm (SSE)}\;.
\end{eqnarray}
Let us now turn to
the molecular ortho state, which turns out to dominate in the recent RMC
experiment from TRIUMF \cite{TRIUMF}. One obtains for $p_{1/2}=1$:
\begin{eqnarray}
\Gamma^{\rm RMC}_{O}&=&
30.4\times10^{-3}\,s^{-1} \;{\rm (HBChPT)}\nonumber \\
\Gamma^{\rm RMC}_{O}&=&
30.8\times10^{-3}\,s^{-1}\;{\rm (SSE)}\;.
\end{eqnarray}
Due to the triplet dominance in RMC (as opposed to the singlet dominance 
in OMC) $\Gamma^{\rm RMC}_{O}$ is now roughly proportional to $(1-3/4p_{1/2})$ 
which leads to a big sensitivity of the RMC capture rate to the exact
occupation numbers of the relative
molecular
sub-states. For example, a 5\% uncertainty in the
occupation numbers $p_{1/2}=0.95,\;p_{3/2}=0.05$ would lead to a 13\% change in
the
ortho capture rate $\Gamma^{\rm RMC}_{O}\sim 35\times10^{-3}\,s^{-1}$.
We will discuss the implications of this uncertainty when we compare our
results
with the measured photon spectrum from TRIUMF in the next section. For the
total
capture rate in the TRIUMF experiment
\begin{equation}
\Gamma_{\rm RMC}^{H_2}=f_S\;\Gamma_{\rm sing}^{\rm RMC}+f_O\;\Gamma_O^{\rm RMC}+
f_P\;\Gamma_P^{\rm RMC}
\end{equation}
with $f_S=0.061,\,f_O=0.854,\,f_P=0.085$ \cite{TRIUMF} one would obtain
$\Gamma_{\rm RMC}^{TRIUMF}=(34.3\;[34.8]+{\cal O}(1/M_N^2))\times 10^{-3}\,{\rm
s}^{-1}$
in HBChPT [SSE]. This
leads to a relative branching ratio $Q_\gamma=\Gamma_{\rm RMC} / \Gamma_{\rm OMC}$
\begin{eqnarray}
Q^{\rm HBChPT}_\gamma&=&\frac{34.3\times 10^{-3}\,{\rm s}^{-1}}{504\;{\rm s}^{-1}}=
            6.8 \times 10^{-5}+{\cal O}(1/M_N^2))\nonumber\\
Q^{\rm SSE}_\gamma&=&\frac{34.8\times 10^{-3}\,{\rm s}^{-1}}{504\;{\rm
s}^{-1}}=6.9
             \times 10^{-5}+{\cal O}(1/M_N^2))\;.
\end{eqnarray}
Unfortunately the full relative branching ratio is not accessible in
experiment, as
one has to use a severe cut on the photon energies due to strong
backgrounds. In the TRIUMF experiment
only
photons with an energy $\omega>60$MeV were detected. We therefore now move on
to a
discussion on the photon spectrum.

\subsubsection{Photon spectrum}\label{sec:RMCspec}

\begin{figure}[htb]
\centerline{
\epsfysize=3.5in %4.5
\epsffile{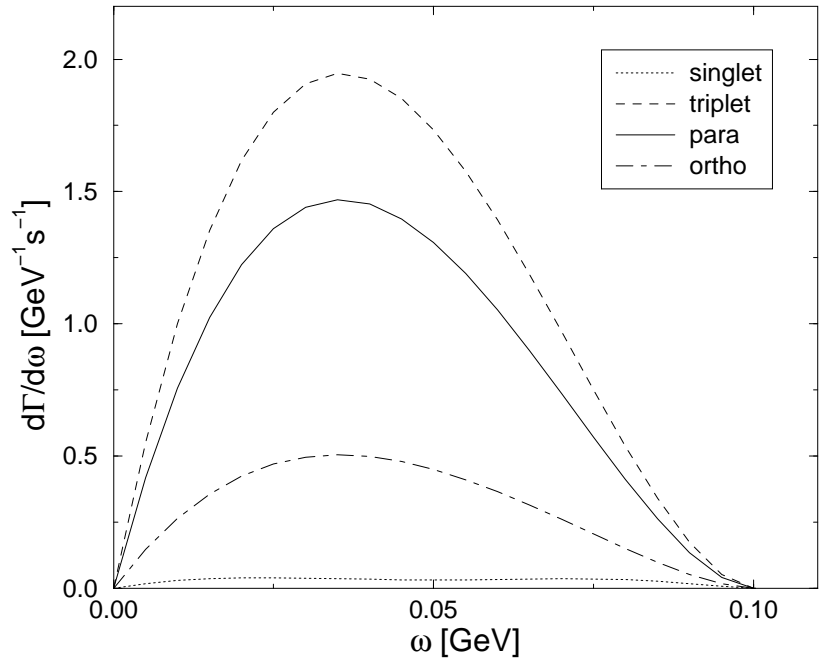}
}
\vskip 1cm %%3

\caption[]{${\cal O}(\epsilon^2)$ SSE
photon spectra for RMC for the singlet, triplet, para (statistical) and
ortho states of the $\mu-p$ system.
}\label{fig:spec}
\end{figure}

\begin{figure}[htb]
\centerline{
\epsfysize=3.5in %4.5
\epsffile{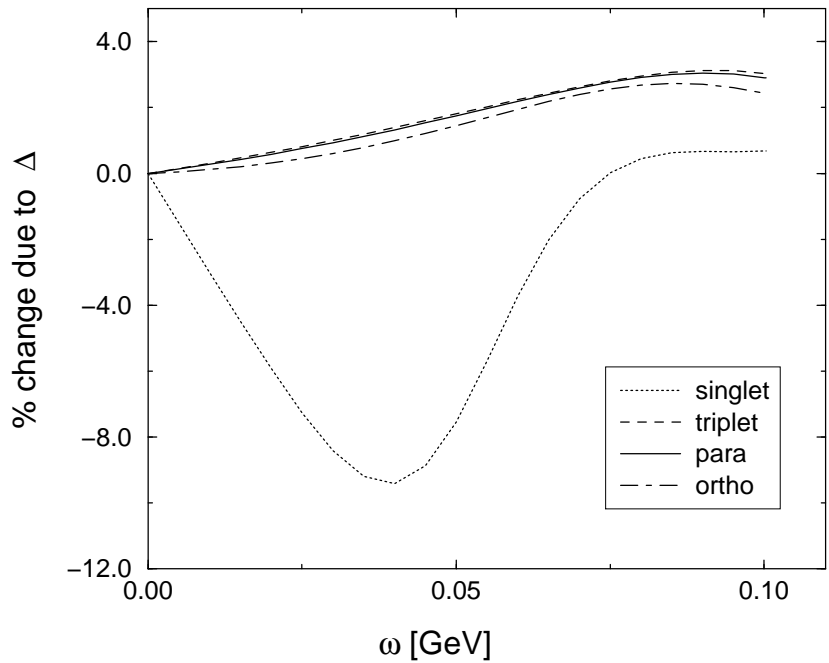}
}
\vskip 1cm %%3

\caption[]{
Relative change in the photon spectra for RMC due to explicit $\Delta$(1232)
related effects
for the singlet, triplet, para (statistical) and
ortho states of the $\mu-p$ system.
}\label{fig:rel}
\end{figure}

The photon spectrum $d\Gamma / d\omega$ can be obtained straightforwardly
from Eq.(\ref{RMCrate}). We refrain from giving the lengthy formulae for
the various atomic states here.
We have calculated the photon spectra to ${\cal O}(p^2)$ in HBChPT and to 
${\cal O}(\epsilon^2)$ in SSE. The resulting curves are very similar. 
In fig.\ref{fig:spec}
we show the SSE ${\cal O}(\epsilon^2)$ results with the coupling values 
$g_{\pi N\Delta}\times b_1 = 1.05\times 12 = 12.6$ for the singlet, triplet,
para and ortho states. The relative difference between the ${\cal O}(p^2)$ 
HBChPT and the ${\cal O}(\epsilon^2)$ SSE calculation for all states
are shown in fig.\ref{fig:rel}, showing explicitly the small role of 
$\Delta$(1232) in RMC. With the exception
of the small singlet, the $\Delta$ effects amount to less than 5\% for all
photon energies. Only in the case of the singlet, a more pronounced
influence of the $\Delta$ is observed. We note that these results
are very similar to the ones found by Beder and Fearing~\cite{BF1,BF2}
for the spectra and the relative contribution from the spin--3/2
resonance, although their calculation is based on a very different
approach. Even the result for the  singlet is comparable though
not identical to the one of Beder and Fearing. It changes sign for photon
energies
of about 74~MeV. Like for the case of the small triplet in OMC, it
is expected that the small singlet (for RMC) is more sensitive to
$1/M_N$ corrections. We have also increased the coupling $b_1$ to
values of 24 and 60 thus enhancing the $\Delta$(1232) contributions in 
the SSE calculation by a
factor of 2 and 5, respectively.  We find very little sensitivity
to this, e.g. the maximum in the photon spectrum for the para state
changes from 1.47~GeV$^{-1}\,{\rm s}^{-1}$ for $b_1 = 12$
to 1.55~GeV$^{-1}\,{\rm s}^{-1}$ for $b_1 = 60$. Correspondingly, the
total decay rate for the para state is increased by 7\%. The smallness of
the $\Delta$(1232) related effects has two reasons. First, as noted already, 
it only appears at NLO, whereas the RMC process is mainly controled by
the leading order effects, hinting to a good convergence behavior of the
chiral expansions. Second, its sole contribution at that order is in
axial--vector correlator, but not in the other three correlators. This
already leads to a statistical suppression as compared to the pure 
nucleonic contributions.
Third, despite the largeness of the coupling $g_{\pi N\Delta}\times
b_1$ (remember that $b_1$ is related to the dominant $M1$ $N\Delta$
transition \cite{GHKP}), the delta contribution to the VA--correlator is 
still smaller than
the one from the nucleon, which is enhanced by the large isovector magnetic
moment.
The dominant axial and axial--vector contribution comes indeed from the leading
pion pole, which is not affected by the $\Delta$(1232) effects. 

To quantify 
these statements,
let us for a moment consider the ${\cal O}(p^2)$ HBChPT calculation. If one
switches off the complete contribution from the axial correlator,
Eq.(\ref{Acorr}), the singlet rate is enhanced by a factor of
1.7, whereas the triplet is decreased by  a factor of about 170~!
Consequently, we then have $\Gamma_{\rm tot} \simeq 7\times 10^{-3}\,{\rm
s}^{-1}$,
which is an order of magnitude smaller than the value given in the previous
section.
If one on the other hand 
sets $\kappa_s = \kappa_v = 0$ in the V, VV and AV correlators,
the singlet rate increases by a factor of about 2.7
and the triplet rate decreases by a factor of 1.3, leading to a total
rate of  68$\times 10^{-3}\,{\rm s}^{-1}$.

\begin{figure}[htb]
\centerline{
\epsfysize=3.5in %4.5
\epsffile{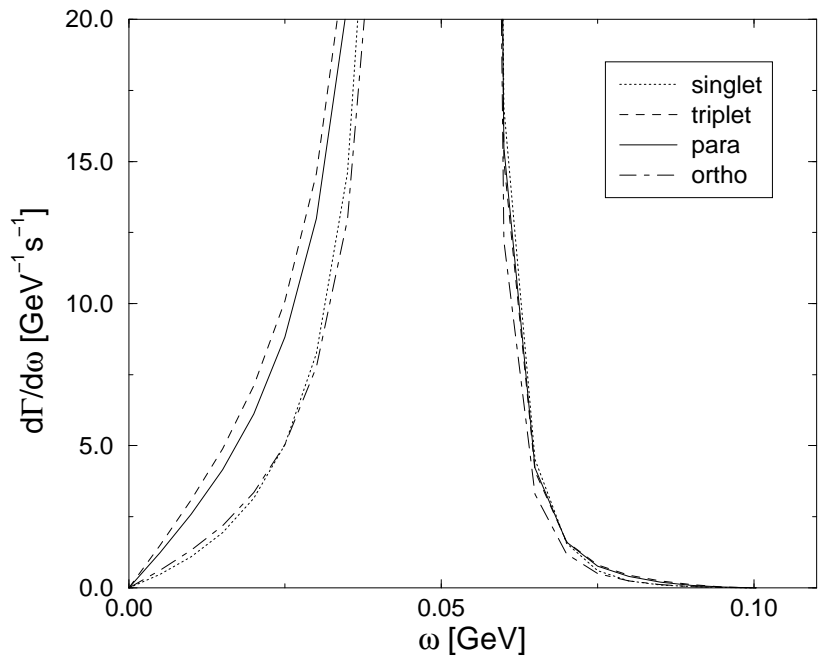}
}
\vskip 1cm %%3

\caption[]{${\cal O}(p^2)$ HBChPT
photon spectra for RMC for the singlet, triplet, para (statistical) and
ortho states of the $\mu-p$ system in the chiral limit.
}\label{fig:specl}
\end{figure}

It is also instructive to consider the chiral limit, $m_\pi =0$.
To be specific, we discuss the ${\cal O}(p^2)$ HBChPT calculation,
with the ${\cal O}(\epsilon^2)$ SSE results being very similar. To 
leading order in $1/M_N$, one
encounters a pole at $\omega = m_\mu /2$ in the A and AV correlators.
This can be seen by looking at pion pole terms in the chiral limit,
which take the form
\begin{equation}
\frac{1}{(l-l')^2} = \frac{1}{m_\mu^2-2m_\mu l_0'}~,
\end{equation}
and using the leading order result $l_0' = m_\mu - \omega + {\cal
  O}(1/M_N)$, cf. Eq.(\ref{l0}). This well-known Bethe--Heitler
pole is independent of the angular variable $x$.
The condition for this pole to appear is that the pion mass
has to be below the muon mass. For zero pion mass, there is another pole
at the same energy for $x= -1$ stemming from the terms $\sim (r-r')^{-1}$.
Furthermore, the chiral limit photon spectra
of the ${\cal O}(p^2)$ HBChPT calculation are
shown in Fig.\ref{fig:specl}. We remark that these spectra
are very different from the ones with the physical pion mass due to
the abovementioned singularities.
We also point out that the pion mass effects are
larger in RMC than in the OMC case discussed above. In particular,
setting e.g. $m_\pi =125\,$MeV, the rate in the para state (total
rate) increases to  99$\times 10^{-3}\,{\rm s}^{-1}$. We will discuss the
implications on the TRIUMF measurement in the following section.

\subsubsection{Discussion of the TRIUMF result for $g_P$}
\label{sec:gPT}

\begin{figure}[htb]
\centerline{
\epsfysize=3.in %4.5
\epsffile{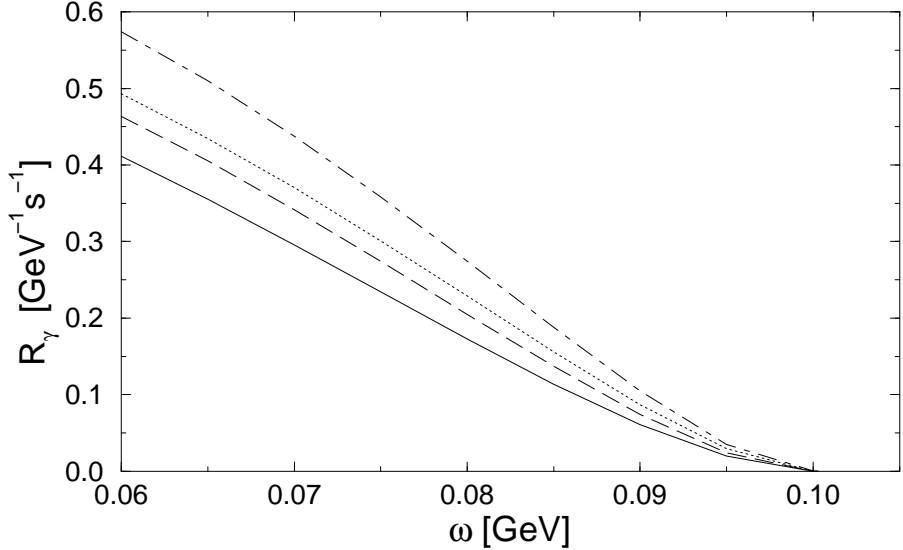}
}
\vskip 1cm %%3

\caption[]{
Photon spectra for RMC for the branching ratios of the singlet, ortho
and para states as used in the TRIUMF analysis. Solid line: Prediction
of the small scale expansion to order $\epsilon^2$. Dash--dotted line:
Same as the solid line but with $g_P$ scaled by a factor
$1.5$. Dashed line: Various small modifications as explained in the
text. Dotted line: Same as the dashed line but using the neutral
instead of the charged pion mass.}\label{fig:specR}
\end{figure}

The photon spectra discussed in section~\ref{sec:RMCspec} allow in
principle to determine the induced pseudoscalar form factor.
The TRIUMF result for $g_P$ is obtained by multiplying the terms
proportional to the pseudoscalar form factor with a constant denoted
$R$ (the momentum dependence assumed to be entirely given by the pion
pole). The value of $R$ is then extracted using the model of Fearing et al.
to match the partial rate for photon energies larger than 60~MeV.
If we perform such a procedure, we get a similar shift in the partial
photon spectra (using the same weight factors for the various $\mu -p$
states as given in ref.\cite{TRIUMF}). It is, however, obvious from
our analysis that such a procedure is not legitimate. By artificially
enhancing the contribution $\sim g_P$ (to simulate this procedure,
we have introduced the factor $R$ in Eqs.(\ref{Acorr},\ref{VAcorr})),
one mocks up a whole class of new contact and other terms not present in the
Born term
model. To demonstrate these points in a more quantitative
fashion, we show in fig.\ref{fig:specR} the partial branching fraction
for our calculation in comparison to the one with $g_P$ enhanced by
a factor 1.5 and a third curve, which is obtained by increasing $g_P$
only by 15\%, take $b_1 =24$ (i.e. enhancing this coupling by a factor
of two) and use $\Delta = 273\,$MeV, since in
the dispersion theoretical analysis of pion--nucleon scattering the
pole in the $P_{33}$ partial wave is located at $W = 1210\,$MeV.
This is shown in fig.\ref{fig:specR} by the dashed line and it shows
that such a combination of small effects can explain most (but not
all) of the shift in the spectrum. This is further sharpened by using
now the neutral pion mass of 134.97~MeV instead of the charged pion
mass, leading to the dotted curve in fig.\ref{fig:specR}. Since the
pion mass difference is almost entirely of electromagnetic origin, one
might speculate that isospin--breaking effects should not be neglected
(as done here and all other existing calculations). Furthermore, as
discussed above a slight change in the occupation numbers $p_{1/2}$
and $p_{3/2}$ would also lead to an increase in the ortho capture rate
which could close the gap between the empirical and theoretical
results. For example the dashed curve in fig.\ref{fig:specR}
would be moved from 0.42 to 0.48
while the dotted one would go from 0.50 to 0.54 with $p_{1/2}=0.95$
and  $p_{3/2}=0.05$.
The situation is reminiscent
of the sigma term analysis, where many small effects combine to give
the sizeable difference between the sigma term at zero momentum
transfer and at the Cheng-Dashen point. We further point out that
although the present investigation combined with the findings
in ref.\cite{AM} does not seem to give any large new term at order
$\epsilon^2$ or from the chiral loops at third order, it can not be excluded
that
one--loop graphs with insertions from the dimension two chiral
Lagrangian (which are formally of fourth order) can give rise to
larger corrections than the third order loop and tree terms calculated
in ref.\cite{AM}. In fact, as we noted before, the photon spectrum is
more sensitive to changes in the anomalous magnetic moments than to
the induced pseudoscalar coupling. Therefore it can be speculated that
one--loop graphs with exactly one insertion $\sim \kappa_v$ can generate large
corrections. This appears plausible but needs to be supported by a
fourth order calculation. Such a mechanism would, however, be much
more natural than the simple  rescaling of $g_P$ based on
tree level diagrams only. Another point against this simple rescaling comes 
from OMC. Indeed if this rescaling holds for RMC it should also hold
for OMC. We thus have performed a similar calculation in OMC. Taking the
same value for R, one would obtain 
\begin{eqnarray}
\Gamma_{\rm OMC}^{R=1.5} = 172.8 \times s^{-1}, \quad  
\Gamma_{\rm OMC}^{\rm sing,\;R=1.5} = 634.6 \times s^{-1}, \quad 
\Gamma_{\rm OMC}^{\rm trip,\;R=1.5} = 18.9 \times s^{-1}, 
 \end{eqnarray}
leading to $\Gamma^{\rm OMC,\;R=1.5}_O = 477 \times s^{-1}$, which is lower than
the error bars of the experimental result from Bardin et al. \cite{bald}. 
As expected, the  
singlet and triplet capture rates are much more sensitive to the details
of the interaction than the total rate. As a conclusion we note 
that the effect of enhancing the capture rates in RMC via setting $R=1.5$ 
leads to a strong reduction of the corresponding OMC rates leading to 
conflicts with the experimentally determined ortho capture rate.

\smallskip\noindent
To summarize this discsussion, we have pointed out that two effects in
particular have to be investigated in more detail:
\begin{itemize}
\item[-] the occupation numbers of the atomic structure in muonic atoms/molecules
need to be carefully re-examined by experts in this field.
\item[-] the N$^2$LO calculation should be redone including all isospin breaking effects
because of the sensitivity to the exact pion mass in the pion-pole 
contributions, for example.
\end{itemize}
\noindent
The sum of these small effects should explain the observed 
photon spectrum, as we believe that the proper hadronic/weak physics part
is well under control by now, as our analysis has
re-confirmed. A simple rescaling of the pseudoscalar coupling constant
should no longer be considered. 

%%%%%%%%%%%%%%%%%%%%%%%%%%%%%%%%%%%%%%%%%%%%%%%%%%%%%%%%%%%%%%%%%%%%%%%%%%
\section{Summary}
In this manuscript, we have considered ordinary and radiative muon capture
on the proton in the framework of the small scale expansion to third and second
order in small momenta, ${\cal O}(\epsilon^{3})$ and ${\cal O}(\epsilon^2)$, respectively.
We have also discussed the induced pseudoscalar form factor of the nucleon
and its determination from the TRIUMF RMC data. The pertinent results of this
investigation
can be summarized as follows:

\begin{itemize}

\item[(i)] To third order in the small scale expansion, ordinary muon
capture is almost not affected by $\Delta$(1232) isobar effects.
The only effect comes via the Pauli radius and is extremely small. The NLO contribution to the total capture rate amounts to a 25\% correction
of the leading term. This is in agreement with naive dimensional counting,
which lets one expect corrections of the size $m_\mu/\Lambda_\chi$.
This calculation involves very few and well-controlled parameters.
We argued that the formula  derived from a relativistic
calculation, see Eq.(\ref{san}), and extensively used in the literature does not hold
in the modern view point of power counting. We have stressed the importance
of the upcoming PSI experiment\cite{proposal}.
\item[(ii)] To second order in the small scale expansion, we have considered 
radiative muon capture.
$\Delta$(1232) related effects only appear at NLO and its effects
on the total capture rate and the photon spectrum are of the order of
a few percent. The smallness of the $\Delta$(1232) contributions is due to a 
combination of
effects as discussed in section~\ref{sec:RMCspec}. This agrees with earlier
findings in a more phenomenological approach~\cite{BF2}. Isobar effects can
therefore not resolve  the discrepancy between the TRIUMF measurement
for the partial decay width $\Gamma (\omega > 60\,$MeV) and the theoretical
predictions. We have, however, pointed out severe loopholes concerning the
extraction
of $g_P$ as done in ref.\cite{TRIUMF}, one of them being the contradiction
with the OMC data. In our opinion the most probable
explanation
of the discrepancy is a combination of many small effects, as detailed
in section~\ref{sec:gPT}.

\item[(iii)] The induced pseudoscalar form factor measured in charged
pion electroproduction is not very well determined, but clearly is in
agreement with the one--loop chiral perturbation theory
prediction~\cite{bkmgp}. A more precise measurement for small invariant
momentum transfer squared is called for~\cite{MAMI}.

\end{itemize}
%%%%%%%%%%%%%%%%%%%%%%%%%%%%%%%%%%%%%%%%%%%%%%%%%%%%%%%%%%%%%%%%%%%%%%%%%%
\section{Acknowledgments}
We would like to thank V. Markushin for useful discussions. We are grateful
to Shung-Ichi Ando, Fred Myhrer and Kuniharu Kubodera for communicating the
results of ref.\cite{AKM} prior to publication.

%%%%%%%%%%%%%%%%%%%%%%%%%%%%%%%%%%%%%%%%%%%%%%%%%%%%%%%%%%%%%%%%%%%%%%%%%%
\appendix
\section{Feynman rules}
\label{app:feyn}
\def\theequation{\Alph{section}.\arabic{equation}}
\setcounter{equation}{0}
In this appendix, we collect the pertinent Feynman rules for calculating
OMC and RMC. These read:

\medskip

\noindent Isovector vector source in ($q_\mu$), nucleon:
\beq
i\,v\cdot {\bf v}+\frac{i}{2 M_0}\left((r_i+r_f)\cdot {\bf v}-(r_i+r_f)\cdot
v\,
v\cdot {\bf v}\right)+\frac{i}{M_0}\,(1+\dot{\kappa}_v)\,[S\cdot {\bf v},S\cdot
q]~,
\eeq
Isovector axial source in, nucleon:
\beq
i\,2\,\dot{g}_A\,S\cdot{\bf a}-i\frac{\dot{g}_A}{M_0}\,S\cdot(r_i+r_f)\,v\cdot
{\bf a}~,
\eeq
Isovector axial source in, pion out ($k_\mu$):
\beq
F_0\,\frac{1}{2}\,{\rm Tr}[\tau^i\,
{\bf a}^\dagger \cdot k + {\bf a} \cdot k \, \tau^i]~,
\eeq
Isovector axial source, vector source, pion:
\beq
-\frac{F_0}{2}\,{\rm Tr}\{ [{\bf
v}_\mu,\tau^i]
\,{\bf a}^\mu+{\bf a}_\mu\,[{\bf v}^\mu,\tau^i] \}~,
\eeq
2 vector sources in, nucleon:
\beq
\frac{i}{2 M_0}\left\{ ( {\bf v}+v^{(s)} )^2-\left[
v\cdot ( {\bf v}+v^{(s)} ) \right]^2-\left(1+\dot{\kappa}_v\right)
[S^\mu,S^\nu] \left[
{\bf v}_\mu ,{\bf v}_\nu \right] \right\}~,
\eeq
vector source and isovector axial source in, nucleon:
\beq
-i\,\frac{\dot{g}_A}{M_0}\left[S\cdot\left({\bf v}+v^{(s)}\right)v\cdot{\bf
a}+v\cdot
{\bf a}\,S\cdot\left({\bf v}+v^{(s)}\right)\right]~,
\eeq
$\Delta_\mu^i$ in, nucleon and vector source ($k_\mu$) out:
\beq
+ i\frac{b_1}{2 M_0}\,{\rm Tr}[
\tau^i(k^\mu{\bf v}^\nu-k^\nu{\bf v}^\mu)]\,S_\nu~,
\eeq
Nucleon in, $\Delta_\mu^i$ and vector source ($k_\mu$) out:
\beq
- i\frac{b_1}{2 M_0}\,{\rm Tr}[
\tau^i(k^\mu{\bf v}^\nu-k^\nu{\bf v}^\mu)]\,S_\nu~,
\eeq
Isovector axial source, nucleon, $\Delta_\mu^i$:
\beq
i\,g_{\pi N\Delta}{\rm Tr}[\tau^i {\bf a}_\mu]~.
\eeq

%%%%%%%%%%%%%%%%%%%%%%%%%%%%%%%%%%%%%%%%%%%%%%%%%%%%%%%%%%%%%%%%%%%%%
%\newpage

\bigskip
\bigskip
\bigskip
\bigskip
\bigskip

%%%%%%%%%%%%%%%%%%tables%%%%%%%%%%%%%%%%%%%%%%%%%%%%%%%%%%%%%%%%%%%%%%%%%
\centerline{ {\Large {\bf Tables}}}

\bigskip

\renewcommand{\arraystretch}{1.3}

\begin{center}

\bigskip

\begin{table}[hbt]
\begin{center}
\begin{tabular}{|c|c|c|c|}
\hline
    Quantity & Symbol & Value & Units \\
    \hline
    Proton mass   & $M_p$      & 938.27321   & MeV \\
    Neutron mass  & $M_n$      & 939.56563   & MeV \\
    Nucleon mass  & $M_N$      & 938.91942   & MeV \\
    Pion mass     & $m_\pi$    & 139.56995   & MeV \\
    Muon mass     & $m_{\mu}$  & 105.658389  & MeV \\
    Axial-vector coupling & $g_A$ & 1.2670   & -- \\
    Pion-Nucleon coupling constant & $g_{\pi NN}$& 13.05& -- \\
    Pion decay constant & $F_\pi$ & 92.42 & MeV \\
    isovector Dirac radius & $ r_1^v$ &$ (0.585)^{1/2 }$& fm \\
    ${\cal O}(\epsilon^3)$ SSE isovector Pauli radius &$r_2^v{\rm (SSE)}$ & 
                           $ (0.61)^{1/2}$ & fm  \\
    ${\cal O}(p^3)$HBChPT isovector Pauli radius &$r_2^v{\rm (HBChPT)}$ & 
                           $ (0.52)^{1/2}$ & fm  \\
    isovector axial radius &$r_A $ & $0.65$ & fm \\
    Fine-structure constant & $\alpha$ & 1/137.0359895 & -- \\
    Fermi constant & $G_F$ & $1.16639 \cdot 10^{-5}$ & GeV$^{-2}$ \\
    CKM matrix element & $V_{ud}$ & 0.9740 &  --  \\
\hline   %%\hline
\end{tabular}
\end{center}
\bigskip
\caption{Values of the various masses, couplings and other constants
used in the text.}
\label{tab:val}
\end{table}

\end{center}

%\begin{table}[hbt]
%\begin{center}
%\begin{tabular}{|l|c|c|c|}
%\hline
%    $b_1$ & $\Gamma_{\rm sing}~[s^{-1}]$ &$\Gamma_{\rm trip}~[s^{-1}]$
%          & $\Gamma_{\rm tot}~[s^{-1}]$ \\
%    \hline
%    0.   & 0.0031   & 0.1119   &  0.08xx \\
%    12.  & 0.0029   & 0.1142   &  0.08xx \\
%\hline   %%\hline
%\end{tabular} \end{center}
%\bigskip
%\caption{Singlet, triplet and total RMC capture rates for the
%theory with and without delta, respectively. The total rate refers
%to the para state, i.e. the statistical mixture.
%}\label{tab:cap}
%\end{table}

\newpage

\centerline{ {\Large {\bf Figures}}}

$\,$\vspace{1cm}
%%%%%%%%%%%%%%%%%%  Fig. 1  %%%%%%%%%%%%%%%%%%%%%%%%%%%%%

$\,$\vspace{3cm}
%%%%%%%%%%%%%%%%%%  Fig. 2  %%%%%%%%%%%%%%%%%%%%%%%%%%%%%

$\,$\vspace{1cm}

%%%%%%%%%%%%%%%%%%  Fig. 3  %%%%%%%%%%%%%%%%%%%%%%%%%%%%%

$\,$\vspace{3cm}
%%%%%%%%%%%%%%%%%%  Fig. Yspec  %%%%%%%%%%%%%%%%%%%%%%%%%%%%%

$\,$\vspace{1cm}
%%%%%%%%%%%%%%%%%%  Fig. relD  %%%%%%%%%%%%%%%%%%%%%%%%%%%%%

$\,$\vspace{.2cm}
%%%%%%%%%%%%%%%%%%  Fig. Yspec - cl  %%%%%%%%%%%%%%%%%%%%%%%%%%%%%

%%%%%%%%%%%%%%%%%%  Fig. X %%%%%%%%%%%%%%%%%%%%%%%%%%%%%

\vskip 1cm

%%%%%%%%%%%%%%%%%%  Fig. branching fraction  %%%%%%%%%%%%%%%%%%%%%%%%%%%%%

\end{document}